\documentclass[a4paper,journal,10pt,twocolumn]{IEEEtran}

\usepackage{amsmath,amsfonts,amsthm}
\usepackage{acronym}
\usepackage{graphicx}
\usepackage{balance}
\usepackage{subfig}
\usepackage{caption}
\usepackage{xcolor}
\usepackage{bigstrut}
\usepackage{todonotes} 

\acrodef{IoT}{Internet of Things}
\acrodef{RFF}{Radio Frequency Fingerprinting}
\acrodef{RF}{Radio Frequency}
\acrodef{DL}{Deep Learning}
\acrodef{ML}{Machine Learning}
\acrodef{CNN}{Convolutional Neural Network}
\acrodef{SDR}{Software Defined Radio}
\acrodef{FPGA}{Field-Programmable Gate Array}
\acrodef{AI}{Artificial Intelligence}
\acrodef{PHY}{Physical}

\newtheorem*{remark}{Remark}

\begin{document}
\title{On the Reliability of Radio Frequency Fingerprinting}

\author{
    \IEEEauthorblockN{Muhammad Irfan\IEEEauthorrefmark{1}, Savio Sciancalepore\IEEEauthorrefmark{2}, Gabriele Oligeri\IEEEauthorrefmark{1}}\\
    \IEEEauthorblockA{\IEEEauthorrefmark{1}Division of Information and Computing Technology, \\College of Science and Engineering, \\Hamad Bin Khalifa University, Qatar Foundation - Doha, Qatar.\\ \{muir45306,goligeri\}@hbku.edu.qa\\
    \IEEEauthorblockA{\IEEEauthorrefmark{2}Eindhoven University of Technology, Eindhoven, Netherlands.\\ s.sciancalepore@tue.nl
}}}

\maketitle

\begin{abstract}
Radio Frequency Fingerprinting (RFF) offers a unique method for identifying devices at the physical (PHY) layer based on their RF emissions due to intrinsic hardware differences. Nevertheless, RFF techniques depend on the ability to extract information from the PHY layer of the radio spectrum by resorting to Software Defined Radios (SDR). Previous works have highlighted the so-called ``Day-After-Tomorrow'' effect, i.e., an intrinsic issue of SDRs leading to a fingerprint mutation following a radio power cycle. In this work, we extend such a study by demonstrating that fingerprint mutations appear every time a new FPGA image is reloaded, i.e., when the SDR initiates a new communication. In this context, we provide an in-depth analysis of the reliability of RFF over multiple FPGA image reloading operations, highlighting its ephemeral and mutational nature. We introduce a methodology for abstracting fingerprint mutations into a graph and provide a theoretical framework for assessing fingerprint reliability. Our results show that the common assumption of considering the RF fingerprint as unique and always persistent is incorrect. By combining real-world measurements, high-performance SDRs, and state-of-the-art deep learning techniques, we experimentally demonstrate that radio devices feature multiple fingerprints that can be clustered according to shared features. Moreover, we show that the RF fingerprint is a time-independent probabilistic phenomenon, which requires the collection of multiple samples to achieve the necessary reliability.
\end{abstract}


\IEEEpeerreviewmaketitle

\section{Introduction}
\label{sec:intro}
\IEEEPARstart{I}{n} the evolving landscape of digital communications, ensuring the security and authenticity of devices has become paramount~\cite{hasan2023review}. With the proliferation of \ac{IoT} devices, smartphones, and other wireless devices, traditional security measures often do not provide the necessary protection against advanced cyber threats~\cite{mao2023_comst}. In this context, \ac{RFF} emerges as a breakthrough solution, offering a non-intrusive, highly accurate method for identifying and authenticating devices at the \ac{PHY} layer of the protocol stack, based on their unique RF emissions~\cite{alhazbi2024_iwcmc},~\cite{irram2022_jnca}. \ac{RFF} is based on the principle that all wireless devices emit a unique signal due to variations in their hardware manufacturing process. These tiny differences, often called hardware imperfections, act as a distinctive ``fingerprint'' that can be used to identify a device, among others~\cite{jagannath2022_comnet}. Unlike traditional security measures that rely on software-based encryption or passwords, \ac{RFF} leverages these inherent hardware characteristics, making it difficult (if not impossible) to replicate or spoof~\cite{papangelo2023commag}. The process of \ac{RFF} involves several key steps. A device's raw signal is initially captured through a dedicated device, i.e., a \ac{SDR}, and pre-processed to extract its unique features, including power, frequency, phase, and modulation characteristics. Signal processing and \ac{DL} algorithms are then used to extract these features and distinguish each device's fingerprint. Once a device's \ac{RFF} is captured and added to the model, it can be identified in subsequent communications, enabling, in principle, real-time authentication and monitoring without additional software and tools or modifications to the device~\cite{oligeri2024sac}. 

Recently, many scientific contributions have experimentally verified the potential and implications of \ac{RFF}. Some papers have also investigated real-world phenomena that can affect the stability and reliability of \ac{RFF} (see Sect.~\ref{sec:related_work} for a complete overview). One of the most well-known phenomena that affect \ac{RFF} is the variability of channel conditions, as identified by the authors in~\cite{shawabka2020_infocom}. Such phenomena have also been used to protect against \ac{RFF}, e.g., by authors in~\cite{irfan2024preventingradiofingerprintingfriendly}, where friendly jamming has been deployed to protect against \ac{RFF}. The authors in~\cite{gu2024_tosn} experimentally verified that the temperature conditions around the device have an impact on the radio fingerprint. Similarly, the authors in~\cite{elmaghbub2024_wisec} discovered and characterized the effect of hardware warm-up on the fingerprint. In our recent conference paper in~\cite{alhazbi2023_acsac}, we experimentally demonstrated that hardware reboot on \acp{SDR} causes changes in the radio fingerprint and requires the adoption of image-based \ac{RFF} models, more robust to such changes. Although the above work provides a stemming example of the complex intertwining of phenomena affecting \ac{RFF}, it mainly focuses on keeping \ac{RFF} reliable in the presence of hardware power-cycle operations. The paper, as well as the entire literature on \ac{RFF}, does not delve further into the specific factors leading to the change of the radio fingerprint across power cycles, and it does not show how such findings affect the overall reliability of RFF in the wild. Moreover, the current literature misses a systematic methodology to model the reliability of RFF over time in the presence of factors possibly affecting its consistency across various measurements.

{\bf Contribution.} In this work, we extend our previous contribution in~\cite{alhazbi2023_acsac} by experimentally investigating the reliability of \ac{RFF} on relevant \acp{SDR}. Through systematic experiments carried out on several Ettus Research X310 \acp{SDR} (widely used in the literature for \ac{RFF}), we demonstrate that \ac{FPGA} image reload operations on \ac{SDR} lead to probabilistic changes in the radio fingerprint of the device, thus impacting device authentication (security) and protection against malicious tracking (privacy). Although current state-of-the-art research refers to the multipath effect as the major problem preventing the actual deployment of \ac{RFF}, in this work, we highlight how \ac{RFF} suffers reliability issues across multiple \ac{FPGA} image reloading operations. To this end, we develop a new methodology that abstracts fingerprint mutations into a graph, where the nodes and edges model the persistence of the fingerprint over \ac{FPGA} image reloads, thus allowing the study of fingerprint reliability. Such a methodology can be generalized and applied independently from the factor(s) affecting the stability of RFF, representing a valuable tool to describe its behavior and accuracy over time. As a relevant result, we show that given a ground truth of measures to describe the fingerprint of a transmitter, on average, a receiver is required to observe at least $24$\% of them to authenticate the transmitter with a probability higher than $0.9$ (median). Through such results, we demonstrate that the radio fingerprint is an ephemeral and mutational phenomenon whose reliability is not guaranteed over time when \acp{SDR} are subject to \ac{FPGA} image reloading.

{\bf Paper organization.} The rest of the paper is organized into the following sections. Section~\ref{sec:related_work} summarizes the literature on the reliability of \ac{RFF};   Section~\ref{sec:background} provides preliminary notion on RFF and SDR. Section~\ref{sec:methodology} details the methodology considered throughout the paper for the measurement campaign and the related data analysis, while Sect.~\ref{sec:fingerprint_persistence} introduces the problem of the mutational nature of the radio fingerprint. Section~\ref{sec:graph_based_analisis} presents the graph-based analysis, while Sect.~\ref{sec:discussion} discusses our results, and, finally, we draw the conclusion in Section~\ref{sec:conclusion}.

\section{Related work}
\label{sec:related_work}

Device identification is an important key factor to enable reliable operations of cyber-physical system such as smart grid and industrial control systems. Device authentication can be carried out by resorting to digital signatures, trading off the overhead of key management, which becomes cumbersome in scenarios involving massive deployment of devices, thus requiring new solutions such as hardware fingerprinting~\cite{sanchez2023methodology} and \ac{RFF}. Early research in \ac{RFF} focused mainly on developing custom feature extraction methods using \ac{ML} and \ac{DL} techniques, as evidenced by several studies~\cite{ding2018specific, merchant2018deep, riyaz2018deep, soltani2020more, shen2021radio}.

A significant limitation of \ac{RFF} systems is their sensitivity to the variability of the wireless channel, which can negatively impact performance. This issue has been highlighted in various studies, e.g. in~\cite{shawabka2020_infocom}, where researchers found that training a DL model on data collected on one day and testing it on data acquired on a different day leads to a significant decrease in classification accuracy. In their experiments, the authors trained three DL models on data from 20 wireless transmitters collected over several days in diverse environments, including a cable setup, an anechoic chamber and an open field. The generated models demonstrated inconsistent performance on different days, indicating an intrinsic inability to generalize effectively over time, environments, or conditions.

Further research supports these findings. For example, the authors in \cite{hanna2022wisig} discovered that training and testing on data collected from different receivers further degrades the model performance. These observations emphasize that \ac{RFF} is influenced by a combination of three factors: the transmitter hardware, the receiver hardware and the communication channel. Similar performance drops have been observed when training and testing on a dataset collected in heterogeneous conditions, e.g., by the authors in~\cite{hamdaoui2022}. The authors attributed the issue to changes in channel conditions; however, none of the contributions provided details on the data collection methodology, i.e., if the radio has been power-cycled or the measurement interrupted.

Another work addressing the reliability of \ac{RFF} over time is the one by the authors in~\cite{elmaghbub2021lora}. The authors investigated the performance of the \ac{RFF} systems when applied to the LoRa technology in various scenarios, including indoor and outdoor environments, wireless and wired setups, several distances, configurations, hardware receivers, and locations. Their results confirmed that testing on different days and using different receivers can significantly affect \ac{RFF} accuracy. In addition, they found that variability in protocol configuration and location further affects classification accuracy.

In this context, we also highlight the findings in our conference paper in~\cite{alhazbi2023_acsac}, where we demonstrated the impact of radio power-cycle on the reliability of \ac{RFF}. As mentioned in the introduction, our previous conference paper focused on designing a methodology capable of keeping RFF consistent across power cycles rather than investigating the factors affecting the stability of the fingerprint.

We summarize the literature discussed above in Tab.~\ref{tab:related}, emphasizing the communication technologies, SDR technologies, communication mediums, center frequencies, input types, \ac{DL} networks and observed phenomena considered by such works. Our investigation confirms that previous literature did not identify FPGA firmware reload operations in \acp{SDR} as an element affecting the reliability of RFF, confirming the novelty of these contributions. As a consequence, the current literature misses a systematic methodology allowing us to investigate the impact of FPGA firmware reload operations on the reliability of RFF, possibly generalizable to any physical phenomena affecting RFF. We address the mentioned gaps through this paper.

 \begin{table*}[htbp]
   \centering
   \caption{Related work comparison}
   \label{tab:related}%
   \resizebox{\textwidth}{!}{
     \begin{tabular}{p{2.445em}p{8.22em}p{7.555em}p{8.72em}p{7.39em}p{6.835em}p{8.61em}p{9.72em}p{6.5em}}
     \hline 
     \textbf{Ref.} & \textbf{Communication Technology} & \textbf{SDR Technology} & \textbf{Communication Medium} & \textbf{Center Frequency} & \textbf{Measurement Type} & \textbf{Observed Phenomenon} & \textbf{Input Type} & \textbf{DL Network} \bigstrut\\
     \hline \hline
     \cite{elmaghbub2024_wisec} & IEEE802.11b WiFi packets & USRP B210 & Cable and Wireless & 2.412 GHz  & Training and testing data interleaved by warm up and satabilization of hardware & Impact of warm up and stabilization of hardware on RFF & 128x2 raw IQ samples for Device Classification, 2x17750 raw-IQ samples for Impairment classifcation & Self-designed CNN \bigstrut\\
     \hline
     \cite{ding2018specific} & PHY-QPSK & USRP B210/N210/E310 & Wireless & 800 MHz & Training and Testing data is captured in one shot & Impact of channel conditions & Bispectrum & Self-designed CNN \bigstrut\\
     \hline
     \cite{merchant2018deep} & Zigbee & Rohde \& Schwarz FSW67 & Cable and Wireless & 2.405 GHz & Training and Testing data is captured in one shot & Impact of channel conditions & 1024x2 raw IQ & Self-designed CNN \bigstrut\\
     \hline
     \cite{riyaz2018deep} & IEEE802.11ac & USRP B210 & Wireless & 2.45 GHz & Training and Testing data is captured in one shot & Impact of channel conditions & 2x128 raw IQ & customized Alexnet \bigstrut\\
     \hline
     \cite{shen2021radio} & LoRa  & USRP N210 & Wireless & 868.1 MHz & Training and testing data collected on cross day & Carrier frequency offset (CFO) impact & 256x63 spectrogram & Self-designed CNN \bigstrut\\
     \hline
     \cite{soltani2020more} & PHY-QPSK & - & Wireless & -     & Simulation & Impact of channel conditions & Raw IQ & Self-designed CNN \bigstrut\\
     \hline
     \cite{shawabka2020_infocom} & PHY-BPSK & USRP N210 / X310 & Wireless &  2.4 GHz or 5.8 GHz & Training and testing data collected on cross day & Impact of measurement time & Raw IQ & Customized ResNet50 \bigstrut\\
     \hline
     \cite{al2021deeplora} & LoRa  & USRP N210 & Wireless & 915 MHz & Training and testing data collected on cross day & Impact of measurement time & Raw IQ & Self-designed LSTM/CNN \bigstrut\\
     \hline
     \cite{alhazbi2023_acsac} & PHY-BPSK & X310  & Cable and Wireless & 900 MHz & Training and testing data interleaved by power cycle & Impact of channel conditions and power cycle & 224x224x3 image, Nx2x1 and Nx1x1 Raw-IQ & Resnet50 \bigstrut\\
     \hline
     This paper   & PHY-BPSK & X310  & Cable & 900 MHz & Training and Testing data interleaved with FPGA firmware reload of transmitter X310 & Impact of FPGA firmware reload & 224x224x3 image & ResNet18, AlexNet, Inceptionv3, and SqueezeNet \bigstrut\\
     \hline
     \end{tabular}%
   }
 \end{table*}%

\section{Background}
\label{sec:background}
In this section, we lay the foundation for our discussion on the reliability of \ac{RFF}. We first introduce in Sect.~\ref{sec:rff} how \ac{RFF} can enhance {\em authentication} and {\em non-repudation} of wireless devices by exploiting \ac{PHY} layer information. Then, we discuss how \ac{RFF} is inherently tight to the ability to collect \ac{PHY} layer information from the radio spectrum, thus requiring (in the vast majority of cases) special hardware, i.e., \ac{SDR}, as introduced in Sect.~\ref{sec:sdr}. 

\subsection{Radio Frequency Fingerprinting}
\label{sec:rff}

\begin{figure}
    \centering
    \includegraphics[width=\columnwidth,angle=0,trim = 10mm 40mm 30mm 0mm]{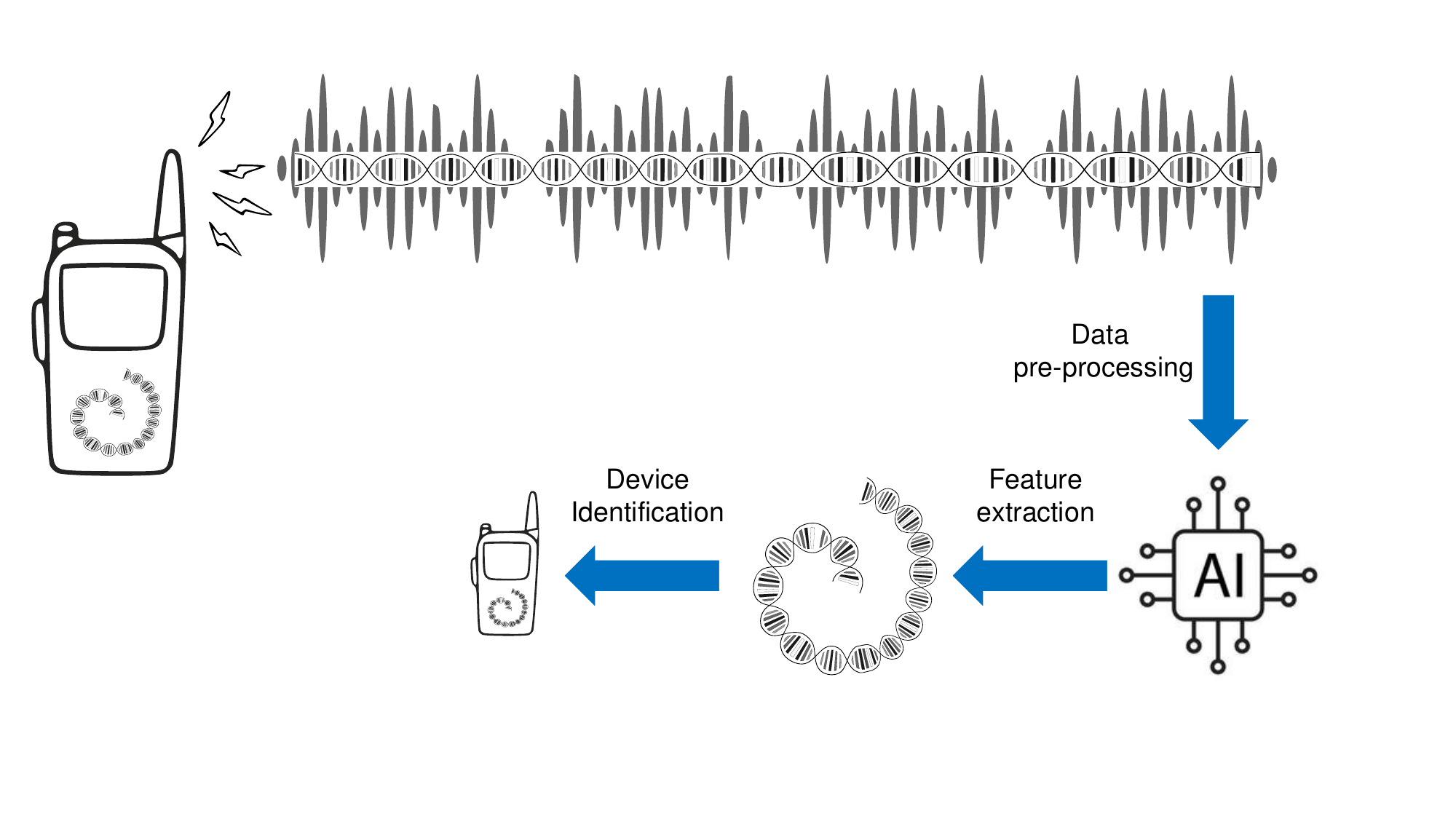}
    \caption{\ac{RFF} overview: the receiver can infer the transmitting source by just eavesdropping a raw signal through a \ac{SDR}. Differences in electronics and software configuration reflect in the over-the-air signal (features), and the receiver can detect and extract these features, thus identifying the signal source.} 
    \label{fig:rff}
\end{figure}

\ac{RFF} is a technique that can enhance the security properties of a system. \ac{RFF} can bolster several security properties, particularly authenticity and non-repudiation, within wireless communications and device identification. Authenticity is a security property that ensures that entities (users, systems, or information) are who they claim to be~\cite{stallings2023}. \ac{RFF} enhances authenticity by providing a means to uniquely identify and verify the identity of wireless devices based on their inherent \ac{RF} emissions. As shown in Fig.~\ref{fig:rff}, since these emissions are unique to each device due to the manufacturing process and are difficult (if not impossible) to replicate or spoof, \ac{RFF} systems process them through \ac{AI} tools to extract features to authenticate devices with overwhelming confidence, ensuring that the device communicating on the network is the one it claims to be~\cite{jagannath2022_comnet}. 
In contrast, non-repudiation is a security property that prevents an entity from denying the authenticity of its signature on a document or a message it sent~\cite{stallings2023}. Although \ac{RFF} is not directly related to digital signatures, it can support non-repudiation in wireless communications by providing undeniable evidence that a specific \ac{RF} device was involved in the communication. Indeed, the unique \ac{RF} fingerprint associated with the device can serve as a form of ``signature'', linking the device to specific actions or communications. \ac{RFF} is used to achieve or improve these properties rather than being a security property itself. It is critical in environments where traditional authentication methods are insufficient or where additional layers of security are desired. For example, in IoT networks, where devices may have limited computational power and cannot handle complex cryptographic operations, \ac{RFF} offers a lightweight yet effective authentication solution.

{\bf RFF and Privacy.} The essence of \ac{RFF}, i.e., identifying devices without requiring active participation or consent, poses inherent privacy concerns. Through RFF systems, devices can be tracked and monitored without the knowledge or approval of the device owner, potentially leading to unauthorized surveillance and data collection. This capability could be exploited for benign purposes, e.g., network security, but also for malicious purposes, i.e., invasive tracking and profiling of individuals based on their devices. This process involves the creation of detailed profiles that, while not directly linked to personal identities, could potentially be correlated with other data sources to deanonymize individuals. The storage and processing of RF fingerprints also require robust data protection measures to prevent unauthorized access and misuse, raising questions about the suitability of existing privacy laws and regulations to protect against such risks. \\
We highlight that this paper aims to investigate the reliability of RFF, especially in the presence of factors like the FPGA image reload on \acp{SDR} part of the system. However, we believe it is worth noting that RFF could be used both as a network security feature and as an attack against the device. Thus, the change of the device fingerprint due to the phenomena investigated in this work can be contextualized for both applications.

\subsection{Software Defined Radios}
\label{sec:sdr}

A \acl{SDR} is a communication system where components that have typically been implemented in hardware (such as mixers, filters, amplifiers, modulators, and demodulators) are instead implemented using software components~\cite{ulversoy2010_comst}. An \ac{SDR} mainly comprises two main components: an RF daughterboard and a \ac{FPGA}. The RF daughterboard performs the down/up conversion of the signal between the baseband and the radio frequency. In contrast, the \ac{FPGA} includes the logic for handling analog-to-digital (ADC) / digital-to-analog (DAC) operations, managing data streams, performing real-time signal processing, and controlling the flow of data between the radio front-end and the host computer. \acp{SDR} are widely used for performing \ac{RFF} since they give access to a \ac{PHY} layer representation of the received signal, i.e., I-Q samples, as discussed later. It should be noted that while transmitters for RFF can be general-purpose devices, the vast majority of the literature on \ac{RFF} features a \ac{SDR} as a receiver, making \acp{SDR} an essential enabling technology for RFF.

Without loss of generality, in this work, we consider the \ac{SDR} USRP X310~\cite{ettus} provided by Ettus Research for transmitting and receiving signals, in line with a significant portion of the literature on RFF. The USRP X310 features two components: (i) firmware and (ii) \ac{FPGA} image~\cite{uhd_man}. The firmware includes the code that handles Ethernet communication, manages the device power-up sequence, and configures the essential operation of the device. In contrast, the \ac{FPGA} image includes the logic for handling ADC/DAC operations, managing data streams, performing real-time signal processing, and controlling data flow between the radio front-end and the host computer. Every time a new radio communication session is initiated, the \ac{FPGA} image is loaded from the internal memory to the actual FPGA. The {\em \ac{FPGA} reloading} is common to all the \ac{SDR}; the only difference is where the \ac{FPGA} image is stored initially either in the host computer or in the memory onboard the \ac{SDR}. In this work, we prove that \ac{FPGA} image reloading changes the radio fingerprint, thus affecting the \ac{RFF} process. Although we consider \ac{SDR} for both transmitting and receiving, we want to stress that our findings are general---thus applying to the scenarios where \ac{SDR} are considered only at the receiver side. In fact, the data collected on the receiver contain the fingerprint of the whole communication chain, i.e., the features of the transmitter, the receiver, and the channel. Although it is a common assumption to ``fingerprint the transmitter'', the classification performed on the data collected on the receiver side involves the discrimination of the whole communication chain. 

\begin{remark}
    An \ac{FPGA} image reloading operation occurs every time an \ac{SDR} initiates a new transmission or reception. Therefore, an \ac{SDR} is affected by an \ac{FPGA} image reload after a power cycle, or more in general, when a communication is interrupted and a new one is initiated.
\end{remark}

\section{Methodology}
\label{sec:methodology}
\begin{figure}
    \centering
    \includegraphics[width=\columnwidth,angle=0,trim = 0mm 0mm 0mm 0mm]{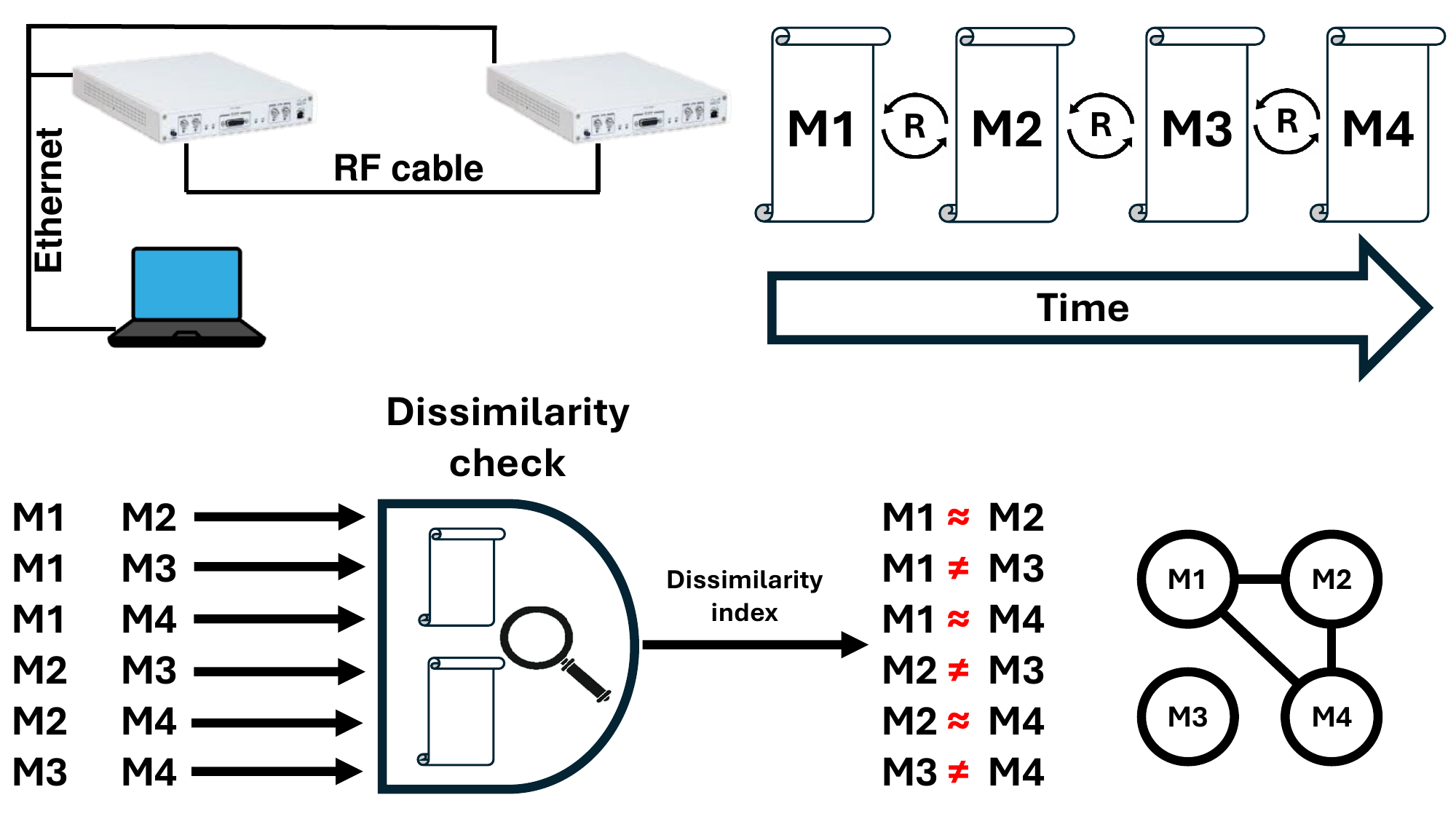}
    \caption{Our methodology involves three phases: (i) Measurement collection, (ii) Dissimilarity analysis of the measurements, and finally, (iii) Analytics extraction.} 
    \label{fig:methodology}
\end{figure}

In this section, we introduce the methodology we adopted to evaluate the reliability of the fingerprint. Figure~\ref{fig:methodology} summarizes our methodology, which involves: (i) collection of consecutive measurements interleaved by \ac{FPGA} image reloading operations, (ii) dissimilarity check analysis on the collected measurements, and finally (iii) the development of a graph-based framework to model fingerprint reliability. We highlight that our measurement campaign considers no radio communication link but only a wired connection. This choice is made to focus our analysis only on the impact of the reload of the \ac{FPGA} image on the fingerprint while mitigating all external factors as much as possible, including multipath fading introduced by the wireless communication channel.

{\bf Measurements.} Our experimental setup takes into account five (5) \acp{SDR}, specifically USRP X310 models equipped with UBX160 daughterboards, in line with recent relevant works on \ac{RFF}~\cite{alhazbi2023_acsac}. In the following, we refer to these radios by the identifiers $\{1, 2, 3, 4, 5\}$. The radios are connected to an HP EliteBook laptop featuring an i7 processor and 32GB of RAM. We consider the scenario in which the communication between the two radios involves an RG58A/U coaxial cable. We also consider GNURadio version 3.8 and design an ad-hoc processing chain for the transmitter and the receiver consistent with the one proposed in previous works~\cite{alhazbi2023ccnc, sadighian2024ccnc, oligeri2024sac}. The center frequency of the radio is 900MHz, the sample rate is 512Kbps, and finally, we considered $0.3$ and $0.1$ as relative transmission power and receiver gain, respectively.

{\bf Data pre-processing.} The data collected through the receiving \ac{SDR} are represented through I-Q samples, which are inherently noisy, even when the transmitter and the receiver are communicating in a controlled environment such as a wired link---being the one considered in this work. To address this, we transform the I-Q samples into images that are more suitable for input into a \ac{CNN}, as acknowledged by several recent works~\cite{alhazbi2023ccnc, sadighian2024ccnc, oligeri2024sac, alhazbi2023_acsac, oligeri2023tifs}. The transformation consists of segmenting the I-Q samples (bi-variate histogram) into a matrix of 224 × 224 tiles, corresponding to the input size (image) required by ResNet18 (and other \ac{CNN} considered in the remainder of this work). Each tile represents a pixel value of a grayscale image and counts how many I-Q samples there are in that region of the I-Q plane.

{\bf Analysis.} We collected 25 measurements for each of the five transmitters consisting of 25 \ac{FPGA} image reloads, one before every new measurement, for a total number of $25\times 5 = 125$~measurements. It is worth mentioning that Fig.~\ref{fig:methodology} only considers a toy example of 4 measurements: $\{M1, \ldots, M4\}$. Moreover, as will become clear in the following, we consider a {\em dissimilarity check} $D(\circ)$ and apply it to all combinations of measurements in groups of 2. We stress that we always compute the dissimilarity check on measurements from the same transmitter-receiver pair. The dissimilarity check returns an index, namely $\delta$, which measures how much the fingerprint is (not) preserved between two measurements $M_x$ and $M_y$, i.e., $\delta = D(M_x, M_y)$, with $x \in \{1, \dots, 25\}$ and $y \in \{1, \dots, 25\}$. Finally, for each transmitter-receiver pair, we make a graph $\mathcal{G} = \{\mathcal{N}, \mathcal{E}\}$, where the nodes $\mathcal{N}$ represent the measurements, and the edges $\mathcal{E}$ represent (an abstraction of) the similarity between two measurements. As an example, in Fig.~\ref{fig:methodology}, the dissimilarity check detects differences between measurements $M_1$ and $M_3$, $M_2$ and $M_3$, and $M_3$ and $M_4$. Therefore, the graph features links between the nodes $M_1$, $M_2$, and $M_4$.

{\bf Dissimilarity check.} We perform a dissimilarity check for each pair of measures in our pool, i.e., $\delta = D(M_x, M_y)$ given a specific transmitter-receiver pair. We consider different \ac{CNN} {\em (ResNet18, AlexNet, Inceptionv3, and SqueezeNet)} featured by Matlab 2023b and split each pair of measures $(x, y)$ into fractions of size 0.6, 0.2, 0.2 for training, validation, and testing, respectively. Furthermore, our methodology involves the conversion of I-Q samples taken from the radio spectrum to images, as already done in other works~\cite{alhazbi2023_acsac, irfan2023isncc, oligeri2023tifs, oligeri2024sac, papangelo2023commag, sadighian2024ccnc, sciancalepore2023jamming}. We highlight that the accuracy computed on the testing set can be considered as the dissimilarity index $\delta$: if \ac{CNN} discriminates between two measurements (high accuracy), the fingerprint is different; conversely, if the accuracy is low, the two measurements are virtually indistinguishable, and thus the fingerprint is the same. Therefore, our final step is the construction of a graph where the nodes represent the measurements and the edges connect the nodes (measures) that are ``similar", i.e., they are characterized by a $\delta$ (dissimilarity index or \ac{CNN} accuracy) which is less than a given threshold ($\tau$). As a toy example, since the output of the \ac{CNN} test is between accuracy values equal to 0.5 (random guess) and 1 (different measurements), we might set $\delta = 0.75$: when the accuracy is less than 0.75, the fingerprint is the same---being the measures indistinguishable---and we draw an edge connecting the two measures; conversely, when the accuracy is higher than 0.75, the fingerprint is different, and the nodes are disconnected. In the remainder of this work, we investigate the trade-off associated with $\delta$ and how this choice affects both the security and privacy of the user.

\section{Fingerprint reliability}
\label{sec:fingerprint_persistence}
\begin{figure}
    \centering
    \subfloat[]{\includegraphics[width=0.5\columnwidth, angle=0, trim = 35mm 90mm 40mm 100mm]{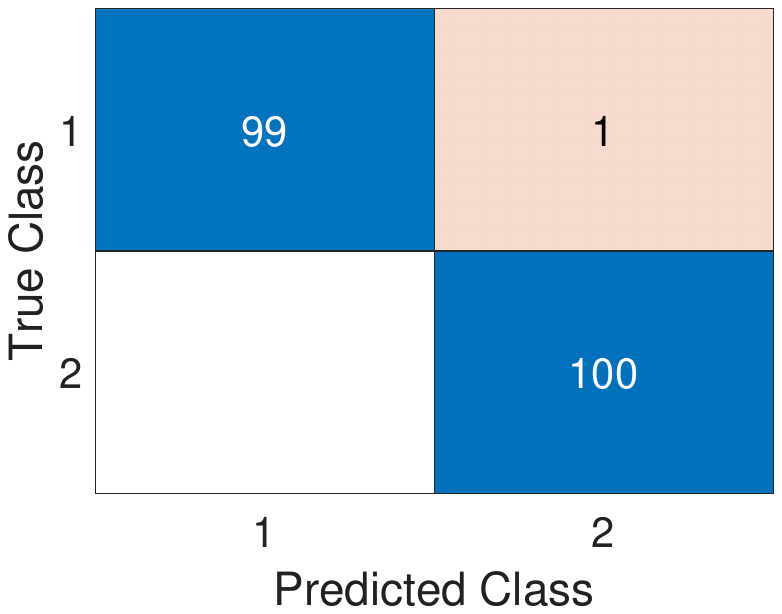}}
    \hfill
    \subfloat[]{\includegraphics[width=0.5\columnwidth, angle=0, trim = 35mm 90mm 40mm 100mm]{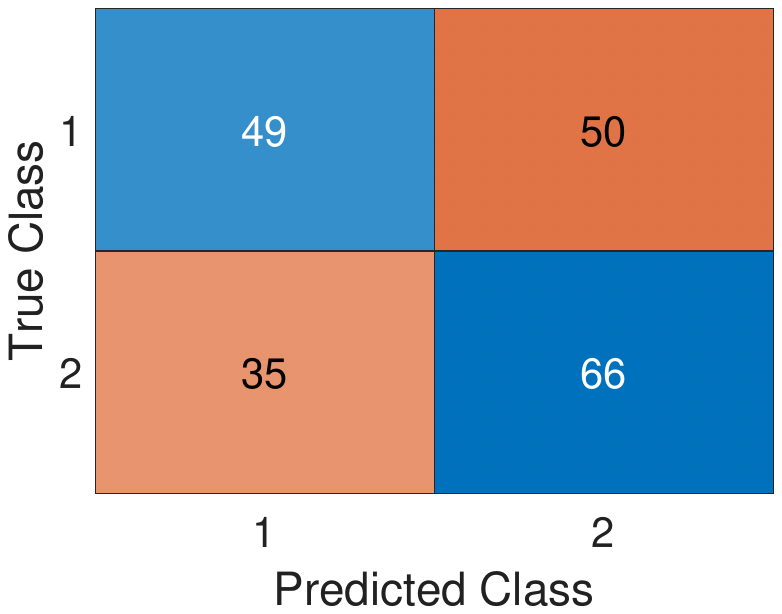}}  
    \caption{We consider two opposite cases: measurements 6 and 11 are characterized by high accuracy (a), while measurements 24 and 25 cannot be distinguished by the classifier (b). We infer that the fingerprint is different for case (a) while it is the same for case (b).}
    \label{fig:cm_4}
\end{figure}

\begin{figure*}
    \centering
    \subfloat[]{\includegraphics[width=0.2\linewidth, angle=0, trim = 30mm 90mm 20mm 30mm]{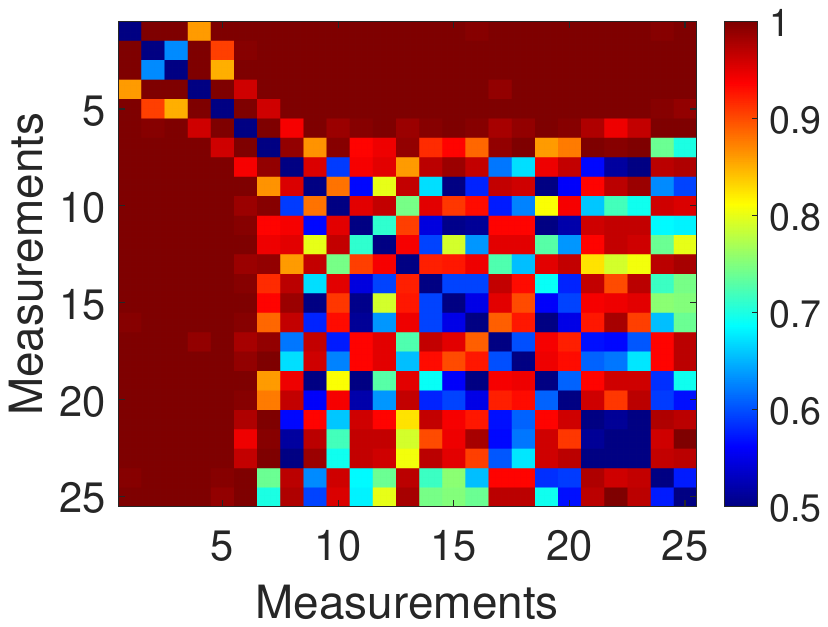}}
    \hfill
    \subfloat[]{\includegraphics[width=0.2\linewidth, angle=0, trim = 30mm 90mm 20mm 30mm]{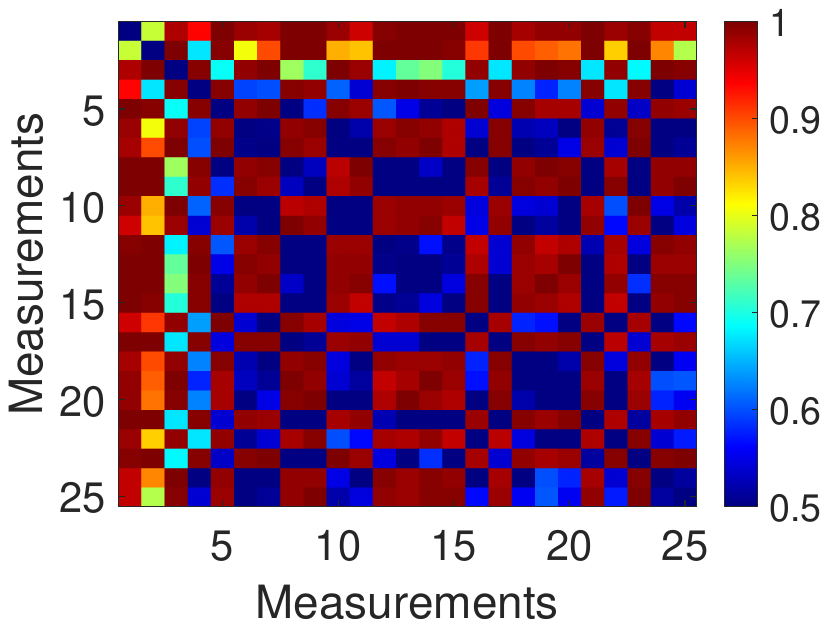}}
    \hfill
    \subfloat[]{\includegraphics[width=0.2\linewidth, angle=0, trim = 30mm 90mm 20mm 30mm]{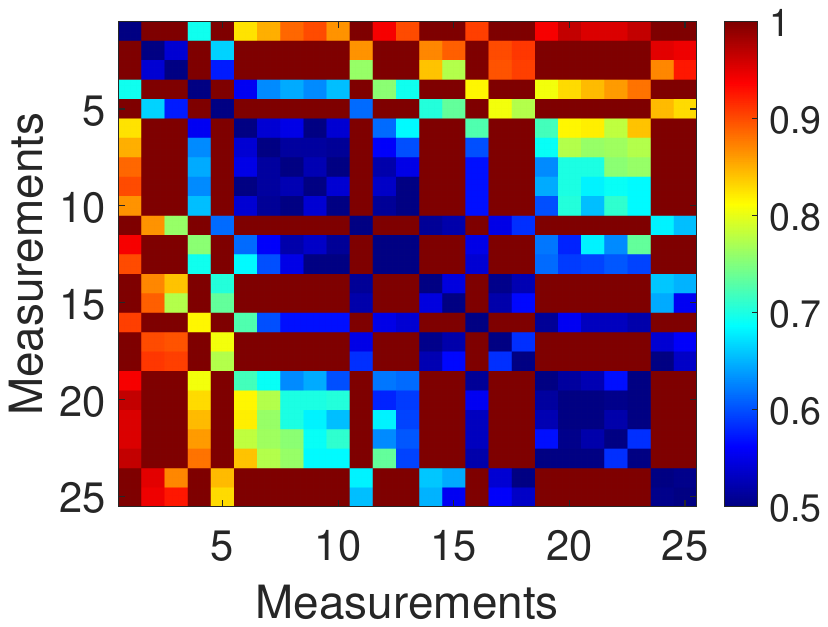}}
    \hfill
    \subfloat[]{\includegraphics[width=0.2\linewidth, angle=0, trim = 30mm 90mm 20mm 30mm]{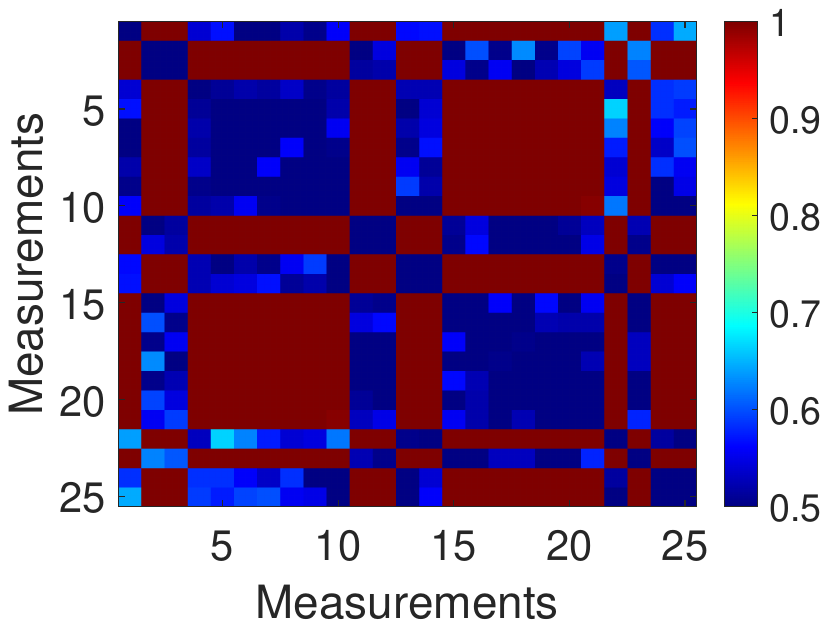}}
    \hfill
    \subfloat[]{\includegraphics[width=0.2\linewidth, angle=0, trim = 30mm 90mm 20mm 30mm]{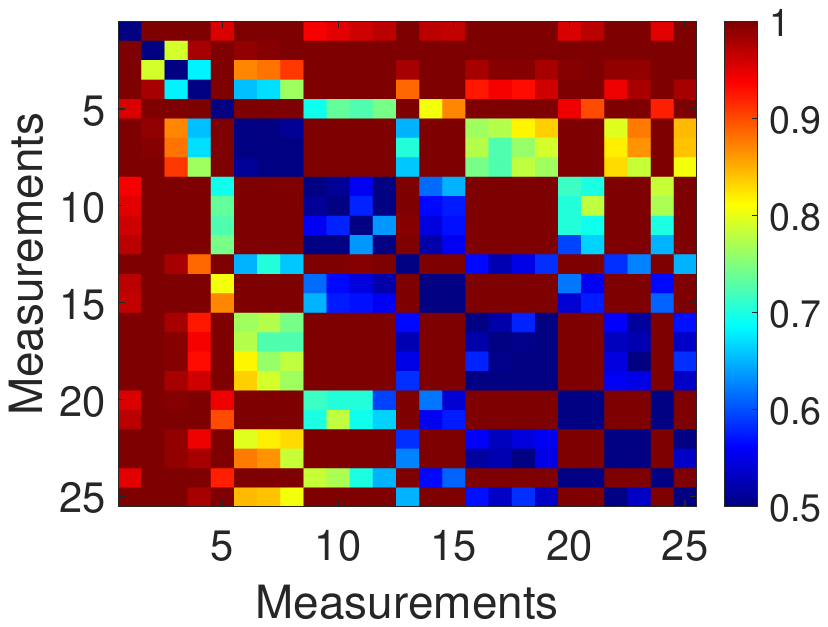}}    
    \caption{Accuracy (dissimilarity index) between consecutive measurements for five transmitters. The color map spans between blue ($\delta = 0.5$) and red ($\delta = 1$).}
    \label{fig:accuracy_R_all}
\end{figure*}

We consider the data set $\mathcal{D}$, which includes 25 measurements interleaved by \ac{FPGA} image reloads for each considered transmitter. For each pair of measurements in $\mathcal{D}$ related to the same transmitter-receiver pair, we compute the dissimilarity check $\delta$ (recall Sect.\ref{sec:methodology} by training, validating and testing data extracted from each pair of measurements. Without loss of generality, we consider the \ac{CNN} ({\em ResNet18}), and we compute the accuracy ($\delta$) for each pair of measurements. This procedure involves ${25 \choose 2} = 300$ binary classifications.  As an example, we consider Transmitter \#1 and two pairs of measurements, i.e., $(6, 11)$ and $(24, 25)$. For each measurement, we consider 300, 100, and 100 images for training, validation, and testing, respectively, where each image has been generated from a sequence of $10^5$ I-Q samples. Figure~\ref{fig:cm_4} shows the confusion matrices associated with the testing performance. The accuracy sums up to 0.995 for Fig.~\ref{fig:cm_4}(a) (only one misprediction out of 200 images). In contrast, Fig.~\ref{fig:cm_4}(b) shows an accuracy of 0.575, i.e., the vast majority of samples (images) from measurements 24 and 25 cannot be distinguished.
As discussed in Sect.~\ref{fig:methodology}, we consider the accuracy of the \ac{CNN} classifier as a dissimilarity index $\delta$. When $\delta \approx 1$, as in the case of Fig.~\ref{fig:cm_4}(a), the classifier can distinguish the measurements, and we claim that the fingerprint of the transmitter is changed from measure 24 to measure 25. In contrast, when $\delta \approx 0.5$, as in Fig.~\ref{fig:cm_4}(b), the classifier cannot distinguish the two measurements, thus we claim that the fingerprint is the same, i.e., the two measurements (6 and 11) are virtually indistinguishable. Given the symmetry of the problem, we assume $D(x, y) = D(y, x) = \delta$ since training, validation, and testing are performed on shuffled fractions of the original dataset, i.e., 0.6, 0.2, 0.2, respectively. Building on such considerations, we map each measurement pair $(x, y)$ to the row-column indexes of a matrix, obtaining Fig.~\ref{fig:accuracy_R_all}. The accuracy $\delta$ is reported through a color map that spans between blue ($\delta = 0.5$) and red ($\delta = 1$).
Figure~\ref{fig:accuracy_R_all} proves that when an \ac{FPGA} image reload is performed (a measurement collection is interrupted and then re-initiated), the radio fingerprint of the tested transmitters exhibits a mutable behavior. First, \emph{none of the transmitters preserves the same fingerprint during all the 25 measurements}. However, the color maps highlight patterns where some measurements have similarities (blue squares), and others differ (red squares). Finally, note that a significant number of measurement pairs have a $\delta$ value close to 0.75: this is the case where the \ac{CNN} does not support (very well) a decision about the presence (or absence) of the fingerprint.

\begin{figure}
    \centering
    \includegraphics[width=\columnwidth,angle = 0,trim = 0mm 0mm 10mm 0mm]{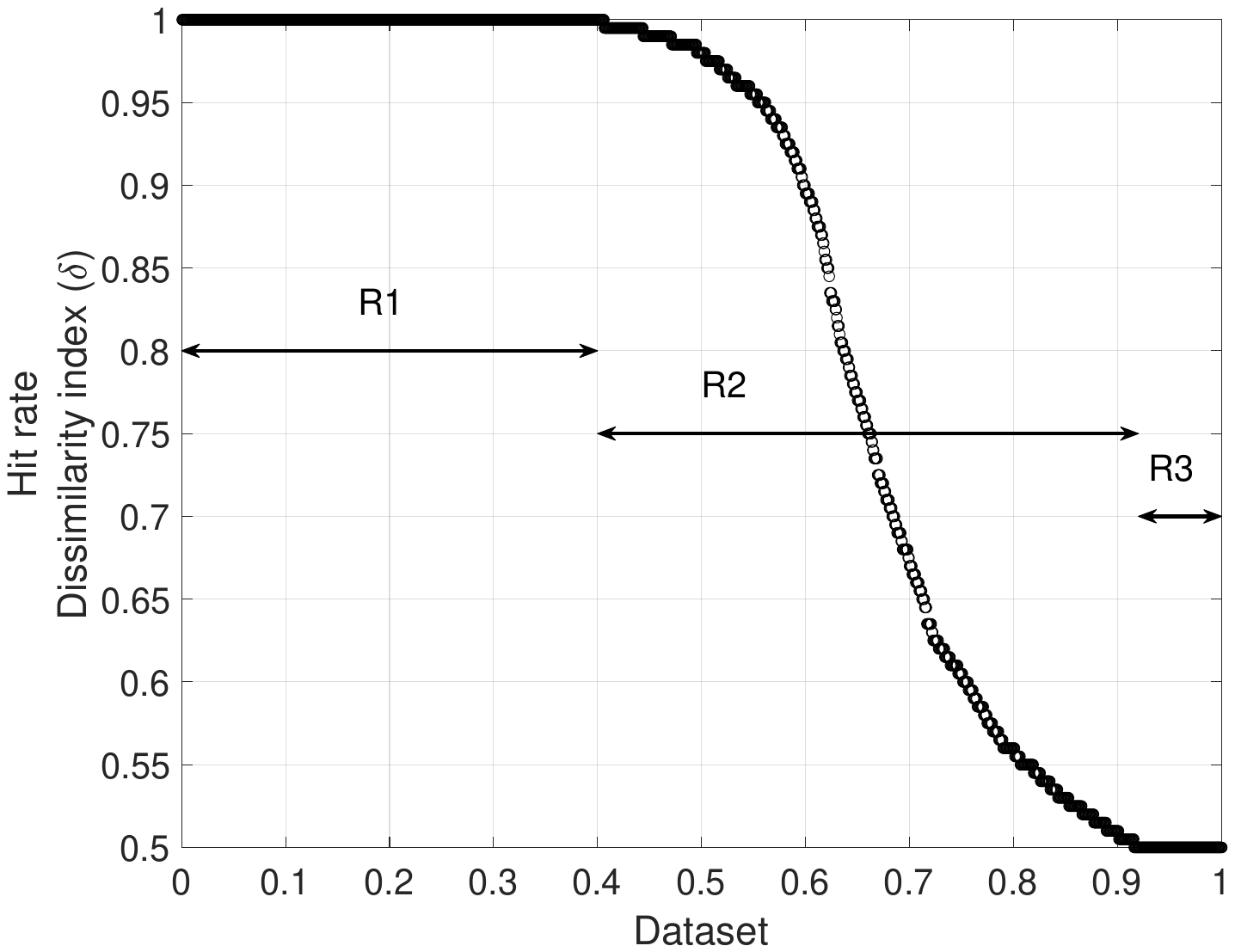}
    \caption{Sorted hit rate (dissimilarity index - $\delta$) while considering all ($300 \cdot 5 = 1500$) measurements. We identify three regions: R1 where $\delta = 1$ (\ac{FPGA} image between measurements is different), R3 where $\delta = 0$ (measurements have the same fingerprint), and finally, R2 where fingerprint detection is challenging.} 
    \label{fig:reloading_all_resnet18_hit_rate}
\end{figure}

\begin{figure}
    \centering
    \includegraphics[width=\columnwidth,angle = 0,trim = 30mm 80mm 40mm 50mm]{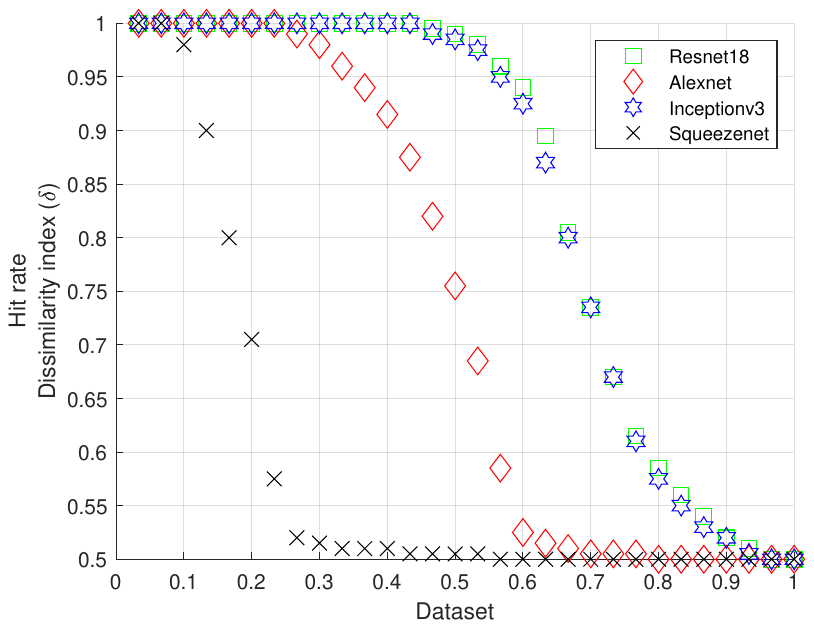}
    \caption{Sorted hit rate (dissimilarity index - $\delta$) while considering all ($300 \cdot 5 = 1500$) measurements and four different \aclp{CNN}: Resnet18, Alexnet, Inceptionv3, and Squeezenet. Resnet18 turns out to be the best trade-off between complexity and performance.} 
    \label{fig:hit_rate_all_networks}
\end{figure}

We now consider all the measurements collected ($300 \cdot 5 = 1500$), independently of the transmitter and the position in the measurement sequence. Figure~\ref{fig:reloading_all_resnet18_hit_rate} shows the sorted hit rate (dissimilarity index $\delta$) as a function of (the fraction of) the collected dataset. We can identify three regions, i.e., R1, where $\delta = 1$, R2, where $1 < \delta < 0.5$, and finally R3 where $\delta = 0.5$. We recall that the common assumption in the literature is to assume an operating point inside R3, i.e., $\delta = 0.5$, i.e., all the measurements are indistinguishable from each other (random guess), thus implying that the fingerprint is consistent throughout all observations. The results in Fig.~\ref{fig:reloading_all_resnet18_hit_rate} prove that this common assumption does not hold true in practice. In fact, given a set of measurements (1,500) interleaved by \ac{FPGA} image reloads, only 130 have a consistent fingerprint, which sums up to about 9\% of the entire data set. In contrast, 600 out of 1,500 (40\%) have a completely different fingerprint (region R1). We also highlight the existence of the region R2 (770 measurements out of 1,500), representing about 51\% of the dataset, where the fingerprint's existence (or absence) cannot be ``clearly'' stated. Finally, we stress that {\em the common assumption of considering two measurements as consistent (same fingerprint for consecutive measurements) when interleaved by a \ac{FPGA} image reloading is wrong in about 91\% of the cases}. 
We now consider different \acp{CNN} to evaluate the relation between the size of the regions \{R1, R2, R3\} and the classifier configuration. Figure~\ref{fig:hit_rate_all_networks} shows the dissimilarity index ($\delta)$ as a function of four different networks, i.e., Resnet18 (recall Fig.~\ref{fig:reloading_all_resnet18_hit_rate} as a reference), Alexnet, Inceptionv3, and Squeezenet. Networks have various degrees of complexity with respect to layers, depth, size, and number of parameters~\cite{irfan2024preventingradiofingerprintingfriendly}. Squeezenet is the simplest network in the \ac{CNN} pool provided by Matlab2023b, while Inceptionv3 is among the most complex. We observe no difference between Resnet18 and Inceptionv3, which confirms that Resnet18 is a good trade-off between complexity and performance. In contrast, Alexnet is characterized by smaller R1 and larger R3 regions. These features confirm that the performance of Alexnet is worse than that of the other networks: the fraction of measurements classified as different drops from about 0.5 to 0.25, while the fraction of similar measurements increases from about 0.1 to 0.35. This result should not lead to the conclusion that a larger fraction of the dataset preserves the fingerprint. In fact, Alexnet performs worse at \ac{RFF}~\cite{irfan2024preventingradiofingerprintingfriendly}. This correlation is also confirmed when looking at the performance of Squeezenet---the network with the worst performance that can be used to solve the classification problem of \ac{RFF}~\cite{irfan2024preventingradiofingerprintingfriendly}. The size of the R3 region depends on two factors: (i) the persistence of the fingerprint and (ii) the inability of \ac{CNN} to solve the \ac{RFF} classification problem. Since we can address the second one considering a better performing \ac{CNN}, such as Resnet18 or Inceptionv3, we can conclude that {\em R3 always represents the most considerable approximation of the fraction of the dataset containing the measurements where the fingerprint can be detected}.

\section{Graph analysis}
\label{sec:graph_based_analisis}
\begin{figure}
    \centering
    \includegraphics[width=\columnwidth,angle = 0,trim = 0mm 0mm 0mm 0mm]{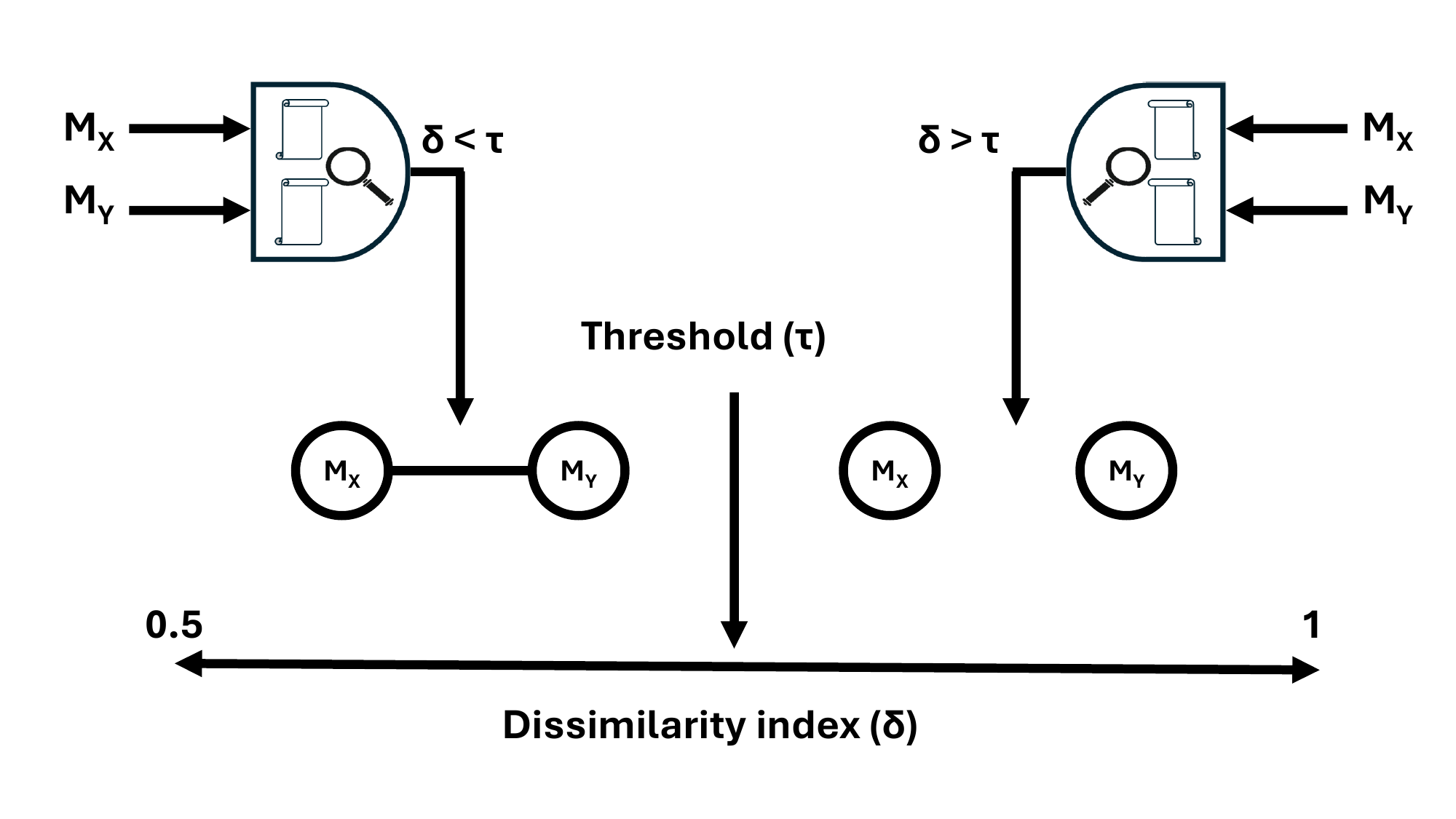}
    \caption{Graph generation from the dissimilarity index: nodes represent measurements, and edges are a function of the dissimilarity index ($\delta$) and a threshold value.} 
    \label{fig:making_edges}
\end{figure}

In this section, we develop a graph-based framework to analyze the mutational nature of the radio fingerprint of a device. As a toy example, we consider the results from Fig.~\ref{fig:reloading_all_resnet18_hit_rate}, associated with the five available transmitters in our pool. In our graph-based model, we represent each measurement through a node, i.e., $\mathcal{N} \in \{1, \ldots, 25\}$. In contrast, edges ($\mathcal{E}$) are a function of the dissimilarity index estimated between two measurements (nodes) when considering Resnet18. In this preliminary example, we assume the existence of an edge between two nodes if the dissimilarity index is less than 0.75, i.e., $\delta < 0.75$. In fact, given two nodes (measurements), if the accuracy returned by Resnet18 is greater than 0.75, we assume that the classifier can discriminate between the two measures. Thus, the fingerprint is different, and no edge connects the two nodes. In contrast, if the accuracy is less than 0.75, we assume that the same fingerprint characterizes the two measurements, and an edge connects the two nodes. Figure~\ref{fig:making_edges} summarizes the methodology described.

\begin{figure*}
    \centering
    \subfloat[]{\includegraphics[width=0.2\linewidth, angle=0, trim = 40mm 90mm 40mm 30mm]{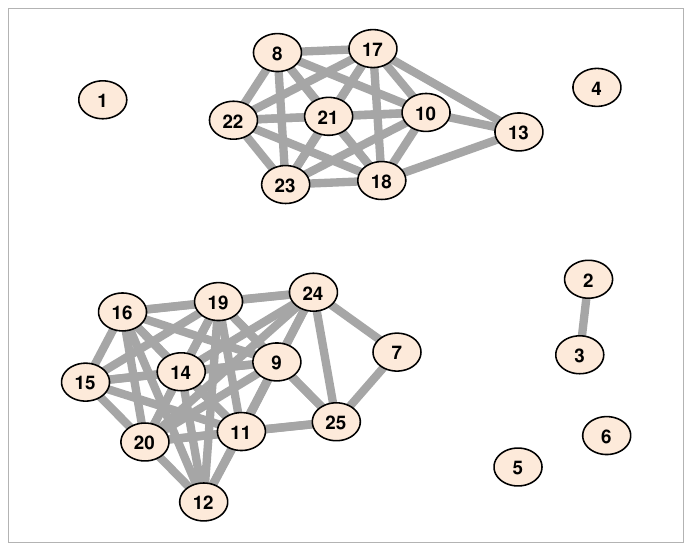}}
    \hfill
    \subfloat[]{\includegraphics[width=0.2\linewidth, angle=0, trim = 40mm 90mm 40mm 30mm]{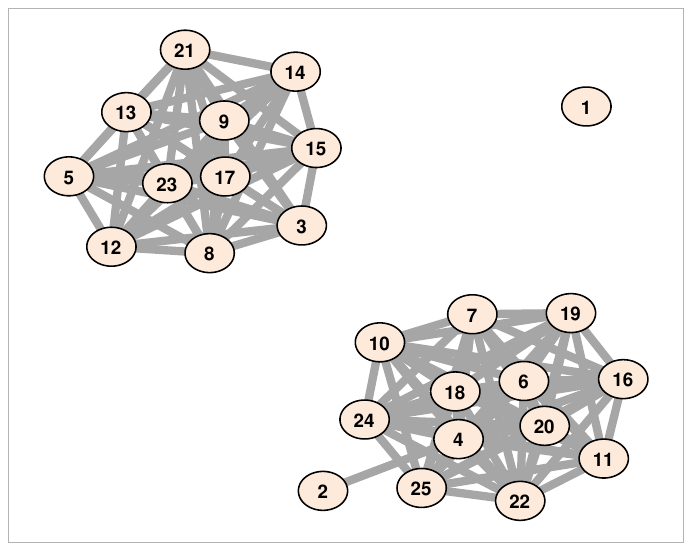}}
    \hfill
    \subfloat[]{\includegraphics[width=0.2\linewidth, angle=0, trim = 40mm 90mm 40mm 30mm]{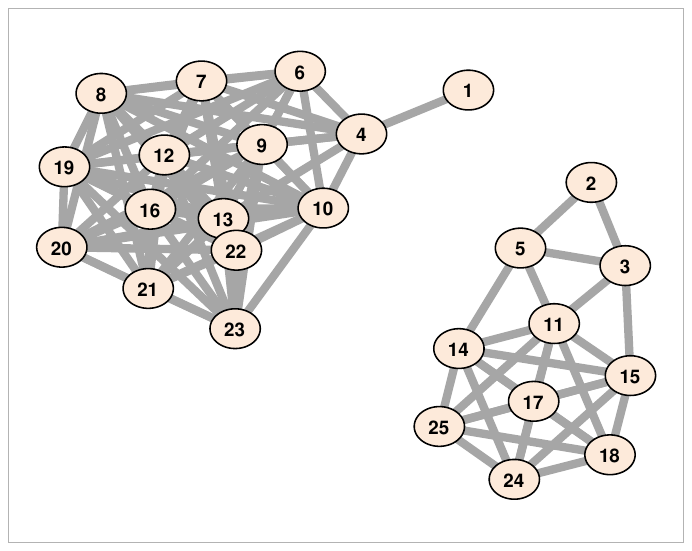}}
    \hfill
    \subfloat[]{\includegraphics[width=0.2\linewidth, angle=0, trim = 40mm 90mm 40mm 30mm]{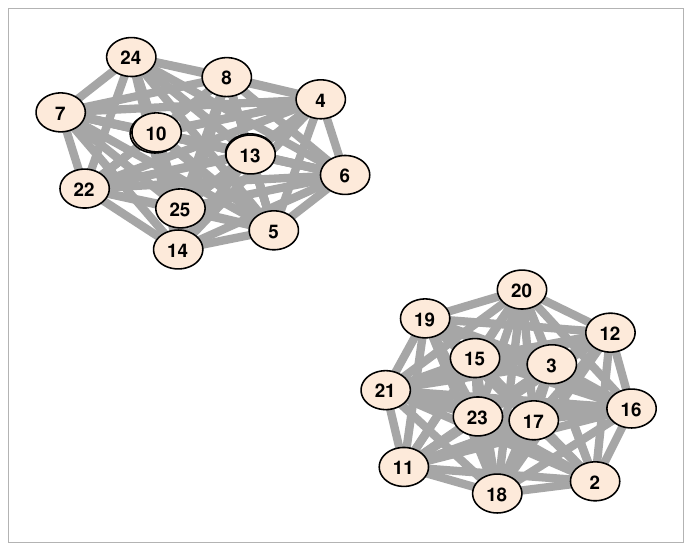}}
    \hfill
    \subfloat[]{\includegraphics[width=0.2\linewidth, angle=0, trim = 40mm 90mm 40mm 30mm]{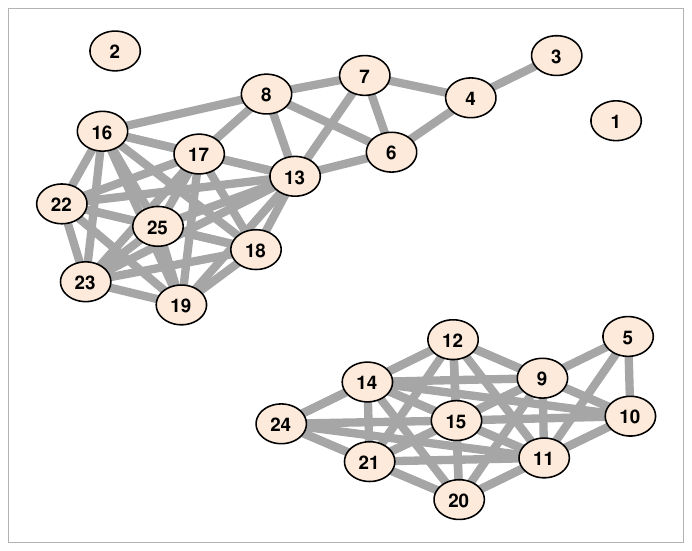}}    
    \caption{Graph-based analysis: measurements that share the same fingerprint are represented through nodes connected by edges.}
    \label{fig:graphs}
\end{figure*}

\begin{figure}
    \centering
    \includegraphics[width=\columnwidth,angle = 0,trim = 30mm 80mm 30mm 90mm]{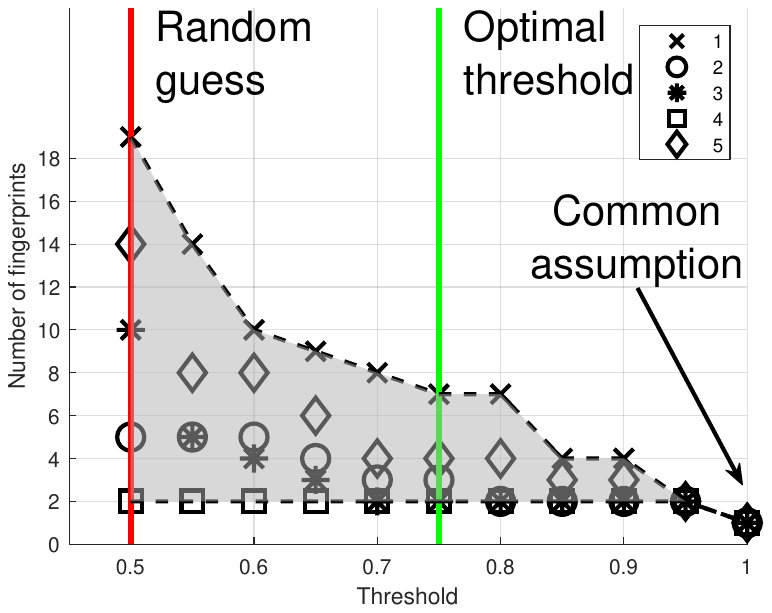}
    \caption{Number of clusters based on the selection of the threshold $\tau$ for all considered transmitters.} 
    \label{fig:states_thr}
\end{figure}

Figure~\ref{fig:graphs} shows the representation associated with the state transitions of the five radios in our pool when considering Resnet18 and a threshold value of 0.75. We first observe that the fingerprint mutates when the \ac{FPGA} image is reloaded, i.e., a measurement is interrupted and subsequently restarted. In fact, the structure of the generated graphs is different. All transmitters in our pool are characterized by two clusters of nodes (measurements) with high similarity ($\delta < 0.75$). Some transmitters (3 out of 5) represented by Fig.~\ref{fig:graphs}(a, b, and e) feature nodes not connected (a, b, and e) or a cluster with only two nodes (a). We highlight that the number of clusters and the degree of the nodes significantly affect the ability to solve the \ac{RFF} problem with a classifier.
On the one hand, the higher the number of clusters featured by a transmitter, the lower the likelihood of collecting a set of measurements with a consistent fingerprint---thus affecting the receiver's ability to identify the transmitter successfully. In fact, under ideal conditions, each transmitter should be characterized by only one cluster. Another critical factor is the degree of the node, i.e., the number of edges each node features. Under ideal conditions, each node belonging to a cluster of $N$ nodes should have a degree of $N$, i.e., the cluster should be fully connected. This situation only occurs for some transmitters considered in our study. For example, Fig.~\ref{fig:graphs}(c) shows that node 1 is only connected to node 4. 
In the remainder of this section, we shed light on the impact of cluster cardinality and node degree on the \ac{RFF} problem.
{\bf Cluster cardinality.} The number of clusters represents the number of internal states the radio transmitter can assume, i.e., the number of different fingerprints. As discussed previously, the number of clusters in the graph depends on the decision (threshold value) whether to assign an edge between two nodes or not, given a dissimilarity index $\delta$. To better understand the phenomena, we analyze the relation between the number of clusters and the threshold value. Figure~\ref{fig:states_thr} shows the number of clusters (different fingerprints) as a function of the threshold value. The common assumption of one single cluster representing one persistent fingerprint throughout all measurements can be achieved by setting the threshold $\tau$ to 1. Indeed, when $\tau = 1$, $\delta \le \tau, \forall \delta$, all measurements are similar (characterized by the same fingerprint) independently of the output of the classifier (recall Fig.~\ref{fig:making_edges}), and therefore the graph is composed of a single fully-connected cluster. Unfortunately, this is not a good approximation of reality.
Transmitter authentication involves the trade-off between hit-and-miss rates. The optimal trade-off is when the threshold is set at the same distance between the random guess (0.5) and the perfect accuracy (1), i.e., a threshold value $\tau = 0.75$. According to such a choice, we make an edge between two nodes of the graph if $\delta < 0.75$, as previously depicted by the example in Fig.~\ref{fig:graphs}. We also observe the (last) extreme case where $\tau = 0.5$. The threshold is set according to the random guess, and therefore, two nodes are connected by an edge on the graph if $\delta <= 0.5$. Of course, this is an extreme case: eventually, two measures are considered to share the same fingerprint if the accuracy is strictly less than 0.5.

Moreover, Fig.~\ref{fig:states_thr} shows that the five transmitters are characterized by different clustering when varying $\tau$. For example, Transmitter \#4 features two clusters almost independently of $\tau$, since the two clusters collapse to 1 only when $\tau = 1$. In contrast, the fingerprints of Transmitter \#1 span between 19 and 1 as a function of $\tau$. Overall, the clustering depends on the distribution of $\delta$ as the classifier's output. Given all the combinations of measurements taken two by two, if the associated dissimilarity indexes $\delta$ are very biased, i.e., either 0.5 or 1, the clustering would be independent of $\tau$. Ideally, this can be represented through a sharp transition (90 degrees) in Fig.~\ref{fig:hit_rate_all_networks}. When the output of the \ac{CNN} gives a $\delta$ value in the range $]0.5, 1[$, the structure of the graph, and in turn the number of clusters, becomes affected by the choice of $\tau$.

Overall, we believe that the number of fingerprints and the fingerprints themselves are fuzzy: the internal state of the transmitter does not jump between entirely different states, but the transition is only partial---thus preserving features from the old state while acquiring new features after the \ac{FPGA} image reload. In fact, given two measurements and an accuracy $\delta = 0.75$---being the output of the classifier---, the dissimilarity index shows that the two measurements are different in 75\% of the cases, i.e., the fingerprint observed during the training phase has changed with respect to the test. In contrast, there is a 25\% fraction of the measurements characterized by false positives and false negatives, showing that the radio preserves the fingerprint (or some of the features) observed during the training time. The internal state of the radio is moving from the old state to the new one.

\begin{figure}[h]
    \centering
    \includegraphics[width=\columnwidth,angle = 0,trim = 30mm 80mm 30mm 90mm]{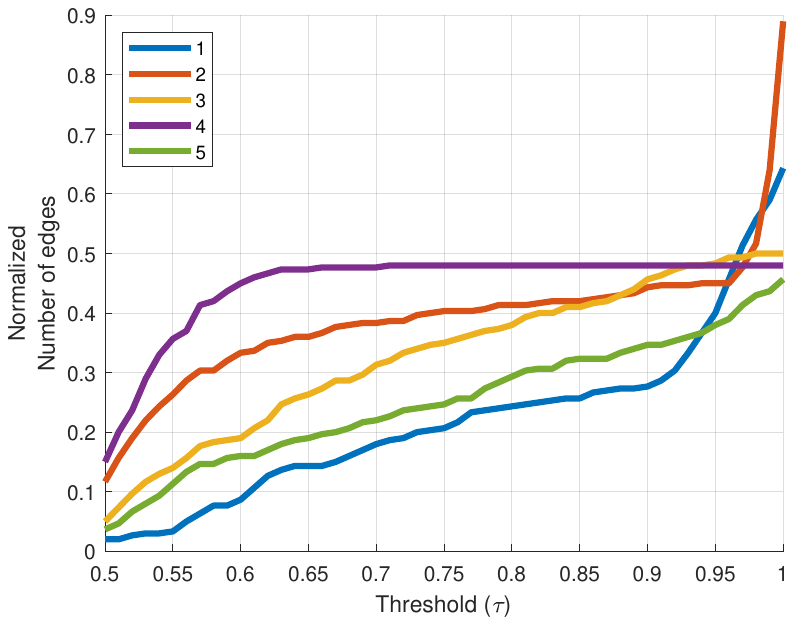}
    \caption{Number of edges (normalized to all the 300 possible ones) between nodes as a function of the threshold $\tau$ for each transmitter in the pool.} 
    \label{fig:edges_threshold}
\end{figure}

{\bf Node degree.} Another essential factor to be considered is how frequently the fingerprint is preserved among the measures, i.e., the number of edges connecting the nodes (the degree of the node). Recalling Fig.~\ref{fig:making_edges}, we observe that two nodes (measures) are linked by an edge (the same or similar fingerprint is detected) when the dissimilarity index is less than a given threshold $\tau$. Therefore, we focus on counting how many edges are in the graph for a given threshold value $\tau$. Figure~\ref{fig:edges_threshold} shows the (normalized) number of edges as a function of the threshold $\tau$ for each of the five transmitters in our pool. The total number of possible edges equals ${25 \choose 2} = 300$. Therefore, we report the fraction of the existing edges with respect to the total number of comparisons made. As a reference case, we discuss $\tau = 1$: two nodes are linked if $\delta < \tau$, i.e., two measures are considered sharing the fingerprint if the accuracy is less than  $\tau = 1$---indeed, if $\delta = 1$, the classifier can distinguish them (with perfect accuracy) and we declare that the two measures are not sharing the fingerprint. Figure~\ref{fig:making_edges} shows that all transmitters are between 0.45 and 0.64, except for Transmitter \#2, which features 89\% 

A common assumption in \ac{RFF} is to consider the identification of a device (in a given pool) as successful when the classifier's performance is higher than a given threshold, e.g., 0.95. Moreover, when the classifier's performance does not meet the given threshold, it is a common assumption to justify it by resorting to the intrinsic variability of the wireless channel (multipath). Our results highlight that this common assumption is wrong. Indeed, the \ac{RFF} problem turns out to be more complex: when $\tau = 0.95$, the number of edges falls in the range $[0.38, 0.49]$ considering all the transmitters in our pool, thus highlighting that two measurements separated by an \ac{FPGA} image reload are likely to have a similar fingerprint in less than 49\% of the cases---being 0.49 the fraction of the nodes (measures) linked by an edge.

\begin{figure}[h]
    \centering
    \includegraphics[width=\columnwidth,angle = 0,trim = 30mm 80mm 30mm 90mm]{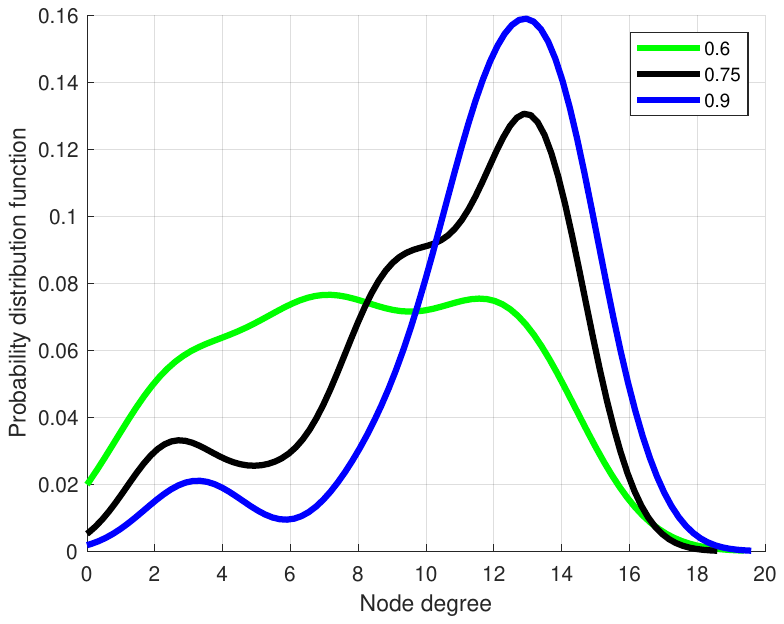}
    \caption{Probability distribution function associated with the degree of the nodes, i.e., number of edges, given $\tau = \{0.6, 0.75, 0.9\}$.} 
    \label{fig:node_degree}
\end{figure}

We also analyze the number of edges per node to highlight how frequently the fingerprint is preserved, given two random measures independently of the transmitter-receiver pair. Figure~\ref{fig:node_degree} shows the probability density function associated with the degree of the node considering all the measures and all the transmitters as a function of the given threshold $\tau$. The threshold value $\tau$ significantly affects the node's degree. Indeed, the probability distribution function changes its shape from almost flat when $\tau = 0.6$ to a peak when $\tau = 0.9$. When the threshold is closer to the random guess ($\tau \approx 0.6$), the number of clusters increases, and therefore, the degree of the nodes depends on the size of the clusters---which become more spread. 

\begin{figure}[h]
    \centering
    \includegraphics[width=\columnwidth,angle = 0,trim = 30mm 90mm 30mm 90mm]{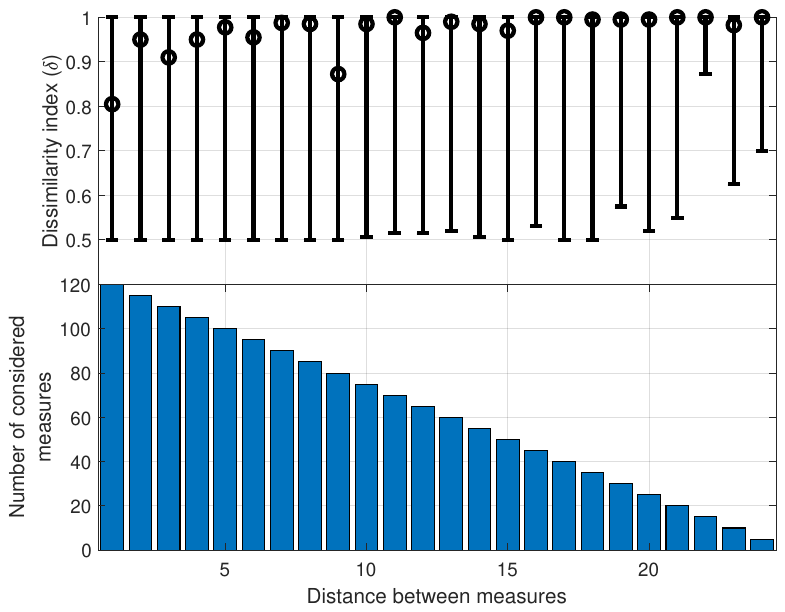}
    \caption{Dissimilarity index ($\delta$) as a function of the (temporal) distance between measurements. The figure highlights how consecutive measurements (distance equal to zero) have the same likelihood of experiencing different fingerprints of randomly chosen measures.} 
    \label{fig:delta_distance}
\end{figure}

Since the measurements have been collected in sequential order, we investigate the persistence of the radio fingerprint over time. Figure~\ref{fig:delta_distance} shows the quantile 5, 50, and 95 associated with the dissimilarity index as a function of the (temporal) distance between measurements, e.g., a distance equal to 1 means that we evaluate the dissimilarity index $\delta$ for all pairs of consecutive measurements considering all transmitters. The results show no relation between the fingerprint and time: two consecutive measurements (distance equal to 1) have the same probability of having the same fingerprint as any other pair of measurements in the dataset. The lower part of Fig.~\ref{fig:delta_distance} highlights the number of considered measures as a function of the defined temporal distance. Indeed, quantiles are computed on different dataset sizes: when distance is 1, the total number of consecutive measures is $120$, i.e., $25$ measures for five transmitters. In contrast, when the distance is 24, the number of pairs is 5, one pair for each considered transmitter.

\begin{figure}[h]
    \centering
    \includegraphics[width=\columnwidth,angle = 0,trim = 30mm 80mm 30mm 90mm]{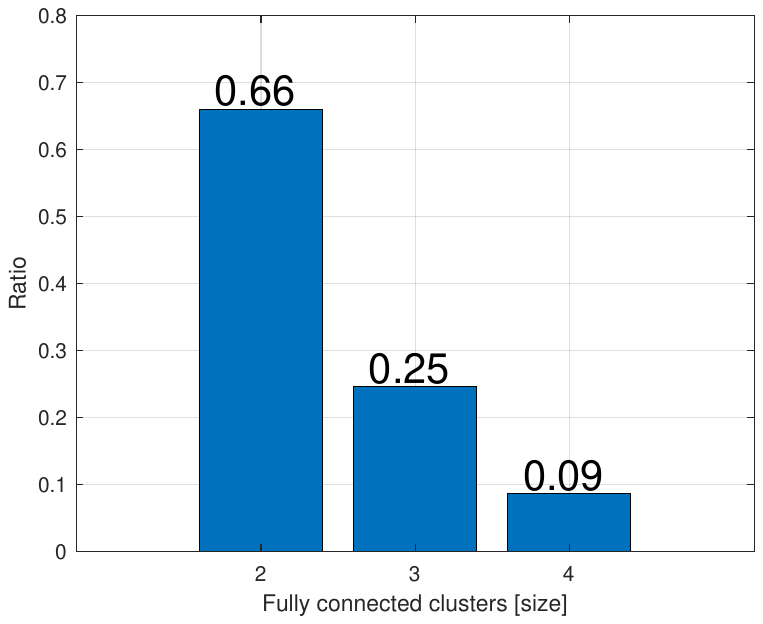}
    \caption{Fraction of the fully connected clusters considering the size of 2, 3, and 4 nodes.} 
    \label{fig:community_size}
\end{figure}

Finally, we perform a more in-depth analysis of the edges connecting the nodes. Since clusters are not fully connected, measurements belonging to the same cluster may not share the same fingerprint. Therefore, we consider all combinations of nodes grouped by 2, 3, and 4, i.e., ${25 \choose 2} = 300$, ${25 \choose 3} = 2300$, and ${25 \choose 4} = 12,650$, respectively. We count how many are fully connected for each configuration and report the ratio. Figure~\ref{fig:community_size} shows the ratio of the fully connected clusters as a function of their size. For example, when considering a cluster composed of 2 nodes, only 66\% 

\section{Discussion}
\label{sec:discussion}
Our study focused on the reliability of \ac{RFF}. Although \ac{RFF} represents an effective solution to authenticate devices based on their unique \ac{RF} emissions, it also comes with significant limitations, emerging from the methodology and hardware required for measurement collection. Our analysis considers data collected from measures interleaved by \ac{FPGA} image reload operations. We stress that this phenomenon is intrinsic to the hardware (\ac{SDR}) required to collect the data from the radio spectrum---thus affecting the vast majority of literature on \ac{RFF}. Although we acknowledge the importance of \ac{RFF} to support device authentication, in this work, we highlight some of its limitations.
On the one hand, these limitations should be addressed to improve device authentication via \ac{RFF}; on the other hand, they can help the end-user avoid being tracked by malicious entities. Our analysis shows that performing \ac{FPGA} image reload operations time by time can be an effective technique to probabilistically randomize the exposure of the device's fingerprint to the radio spectrum. In contrast, the same phenomenon hinders the performance of the authentication process, requiring further investigation. In general, the mutational and transient nature of the radio fingerprint represents a trade-off between security and privacy. Fingerprint mutations hinder at the same time the authentication process, unauthorized identification and tracking of the end-user. We highlight that, at the time of writing, the described factor represents the most efficient way to change a device's fingerprint without involving uncontrollable factors (e.g., the wireless channel) and any other external devices (e.g. a jammer). We translate the problem of assessing the reliability of the \ac{RFF} into a graph-based framework showing how the mutational and transient nature of the fingerprint can be measured in terms of cluster cardinality and degree of nodes. Our methodology can easily be generalized and applied to model the impact of any real-world phenomena affecting RFF, representing a valuable tool to analyze RFF reliability over time. Contrary to common knowledge, our analysis shows that the fingerprint is not just a set of features challenging to detect. The device's fingerprint (probabilistically) mutates every time the FPGA image is reloaded, and the mutation pattern depends on the device itself. 

\begin{figure}
    \centering
    \includegraphics[width=\columnwidth,angle = 0,trim = 30mm 80mm 30mm 90mm]{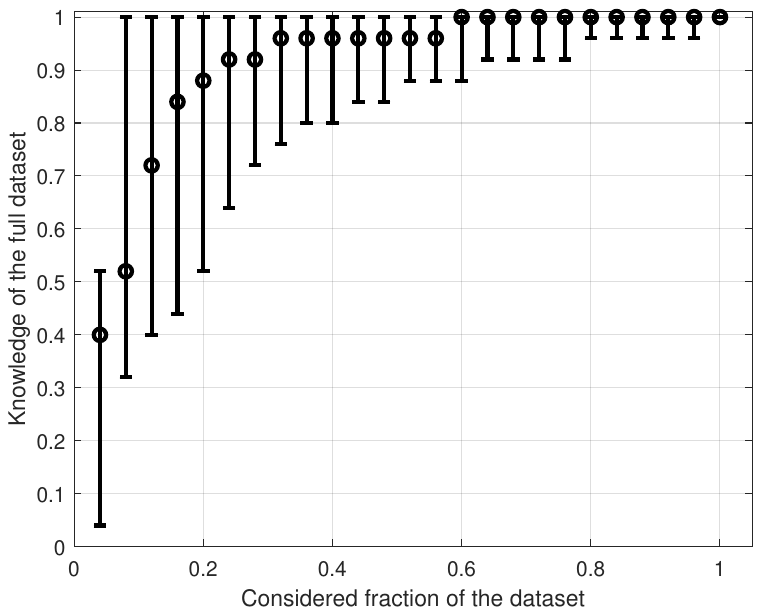}
    \caption{Fingerprint observability: fraction of the whole graph linked to a random set of nodes.} 
    \label{fig:observability}
\end{figure}

Figure~\ref{fig:observability} shows the fraction of the whole graph linked to a random set of nodes. Given a set of randomly taken measures (nodes) associated with a transmitter-receiver pair, we can investigate their connectivity to the rest of the graph, thus inferring the knowledge of the fingerprint phenomenon given a limited number of observations (fraction of considered measurements). 

We assume knowledge of all possible subsets of measures from 1 to 25 spanning from 4\% to 100\% (the entire dataset). Figure~\ref{fig:observability} reports the quantiles 5, 50, and 95 associated with the fraction of the whole graph linked to the subset of nodes considered, independently of the transmitter. For example, we observe that knowledge of one measure exposes 40\% (median) of the phenomenon, while a median higher than 95\% requires at least 15 out of 25, i.e., $60$\% of the dataset. 

{\bf Limitations.} Our study has some limitations that should be acknowledged. We carried out our measurement campaign considering a particular model of \ac{SDR}, i.e., USRP X310 featuring UBX160 daughterboards. We are confident that our results can be extended to other \ac{SDR} models, being all affected by \ac{FPGA} image reloading operations every time a new measurement is started. However, we cannot provide any insight into the behavior of commercial-off-the-shelf devices, thus leaving this investigation to future works. However, we stress that the vast majority of works in the literature involve a \ac{SDR} at least as a receiver, thus being affected by the phenomena discussed in this work.

\section{Conclusion}
\label{sec:conclusion}
In this work, we have introduced a methodology to abstract fingerprint mutations into a graph framework, providing a novel approach to assessing fingerprint consistency and reliability. This approach can help understand the persistence of the fingerprint under different conditions and develop strategies to mitigate the impact of their mutational nature. Based on real-world measures, our experimental results suggest that while \ac{RFF} has excellent potential, its practical deployment requires further research. 

We performed real-world experiments involving multiple FPGA reload operations using \aclp{SDR} connected through wired links, to exclude the impact of the radio channel variability on our observations. 
In contrast to previous results, we have proven that transmitters do not have a consistent fingerprint when FPGA reload operations occur between measurements. As a result, the radio fingerprint is a temporary and mutational phenomenon requiring multiple observations to be fully characterized across FPGA firmware reloads. Moreover, considering a pool of only five transmitters, we observed heterogeneous behaviors in the mutational patterns of the fingerprint. 

Our results have highlighted some common patterns in all the transmitters considered: (i) measures can be grouped into clusters with shared features, (ii) the fraction of measures sharing the same fingerprint is limited, spanning between 66\% and 9\% when considering 2 and 4 measures, respectively, and finally (iii) a complete characterization of the fingerprint phenomenon requires (at least) an observation of 60\% of the collected dataset. Future work will delve into the characterization of the reliability of RFF for commercial-off-the-shelf devices in the wild.


\bibliographystyle{IEEEtran}
\bibliography{main}

\begin{thebibliography}{10}
\providecommand{\url}[1]{#1}
\csname url@samestyle\endcsname
\providecommand{\newblock}{\relax}
\providecommand{\bibinfo}[2]{#2}
\providecommand{\BIBentrySTDinterwordspacing}{\spaceskip=0pt\relax}
\providecommand{\BIBentryALTinterwordstretchfactor}{4}
\providecommand{\BIBentryALTinterwordspacing}{\spaceskip=\fontdimen2\font plus
\BIBentryALTinterwordstretchfactor\fontdimen3\font minus \fontdimen4\font\relax}
\providecommand{\BIBforeignlanguage}[2]{{%
\expandafter\ifx\csname l@#1\endcsname\relax
\typeout{** WARNING: IEEEtran.bst: No hyphenation pattern has been}%
\typeout{** loaded for the language `#1'. Using the pattern for}%
\typeout{** the default language instead.}%
\else
\language=\csname l@#1\endcsname
\fi
#2}}
\providecommand{\BIBdecl}{\relax}
\BIBdecl

\bibitem{hasan2023review}
M.~K. Hasan, A.~A. Habib, Z.~Shukur, F.~Ibrahim, S.~Islam, and M.~A. Razzaque, ``Review on cyber-physical and cyber-security system in smart grid: Standards, protocols, constraints, and recommendations,'' \emph{Journal of network and computer applications}, vol. 209, p. 103540, 2023.

\bibitem{mao2023_comst}
B.~Mao, J.~Liu, Y.~Wu, and N.~Kato, ``{Security and Privacy on 6G Network Edge: A Survey},'' \emph{IEEE Communications Surveys \& Tutorials}, vol.~25, no.~2, pp. 1095--1127, 2023.

\bibitem{alhazbi2024_iwcmc}
S.~Al-Hazbi, A.~Hussain, S.~Sciancalepore, G.~Oligeri, and P.~Papadimitratos, ``{Radio Frequency Fingerprinting via Deep Learning: Challenges and Opportunities},'' in \emph{International Wireless Communications and Mobile Computing (IWCMC)}, 2024, pp. 0824--0829.

\bibitem{irram2022_jnca}
F.~Irram, M.~Ali, M.~Naeem, and S.~Mumtaz, ``{Physical layer security for beyond 5G/6G networks: Emerging technologies and future directions},'' \emph{Journal of Network and Computer Applications}, vol. 206, p. 103431, 2022.

\bibitem{jagannath2022_comnet}
A.~Jagannath, J.~Jagannath, and P.~S. P.~V. Kumar, ``{A comprehensive survey on radio frequency ({RF}) fingerprinting: Traditional approaches, deep learning, and open challenges},'' \emph{Computer Networks}, vol. 219, p. 109455, 2022.

\bibitem{papangelo2023commag}
L.~Papangelo, M.~Pistilli, S.~Sciancalepore, G.~Oligeri, G.~Piro, and G.~Boggia, ``{Adversarial Machine Learning for Image-Based Radio Frequency Fingerprinting: Attacks and Defenses},'' \emph{IEEE Communications Magazine}, pp. 1--7, 2024.

\bibitem{oligeri2024sac}
A.~Sadighian, S.~Sciancalepore, and G.~Oligeri, ``{SatPrint}: Satellite link fingerprinting,'' in \emph{2024 ACM Symposium on Applied Computing (SAC)}, 2024, pp. 177--185.

\bibitem{shawabka2020_infocom}
A.~Al-Shawabka, F.~Restuccia, S.~D’Oro, T.~Jian, B.~C. Rendon, N.~Soltani, J.~Dy, S.~Ioannidis, K.~Chowdhury, and T.~Melodia, ``{Exposing the fingerprint: Dissecting the impact of the wireless channel on radio fingerprinting},'' in \emph{IEEE INFOCOM 2020-IEEE Conference on Computer Communications}.\hskip 1em plus 0.5em minus 0.4em\relax IEEE, 2020, pp. 646--655.

\bibitem{irfan2024preventingradiofingerprintingfriendly}
M.~Irfan, S.~Sciancalepore, and G.~Oligeri, ``{Preventing Radio Fingerprinting through Friendly Jamming},'' \emph{arXiv preprint arXiv:2407.08311}, 2024.

\bibitem{gu2024_tosn}
X.~Gu, W.~Wu, A.~Song, M.~Yang, Z.~Ling, and J.~Luo, ``{RF-TESI: Radio Frequency Fingerprint-based Smartphone Identification under Temperature Variation},'' \emph{ACM Transactions on Sensor Networks}, vol.~20, no.~2, pp. 1--21, 2024.

\bibitem{elmaghbub2024_wisec}
A.~Elmaghbub and B.~Hamdaoui, ``{No Blind Spots: On the Resiliency of Device Fingerprints to Hardware Warm-Up Through Sequential Transfer Learning},'' in \emph{Proceedings of the 17th ACM Conference on Security and Privacy in Wireless and Mobile Networks}, ser. WiSec '24, 2024, p. 134–144.

\bibitem{alhazbi2023_acsac}
S.~Alhazbi, S.~Sciancalepore, and G.~Oligeri, ``The {Day-After-Tomorrow}: On the performance of radio fingerprinting over time,'' in \emph{Proc. of the 39th Annual Computer Security Applications Conference}, 2023, pp. 439--450.

\bibitem{sanchez2023methodology}
P.~M.~S. S{\'a}nchez, J.~M.~J. Valero, A.~H. Celdr{\'a}n, G.~Bovet, M.~G. P{\'e}rez, and G.~M. P{\'e}rez, ``A methodology to identify identical single-board computers based on hardware behavior fingerprinting,'' \emph{Journal of Network and Computer Applications}, vol. 212, p. 103579, 2023.

\bibitem{ding2018specific}
L.~Ding, S.~Wang, F.~Wang, and W.~Zhang, ``Specific emitter identification via convolutional neural networks,'' \emph{IEEE communications letters}, vol.~22, no.~12, pp. 2591--2594, 2018.

\bibitem{merchant2018deep}
K.~Merchant, S.~Revay, G.~Stantchev, and B.~Nousain, ``Deep learning for rf device fingerprinting in cognitive communication networks,'' \emph{IEEE Journal of Selected Topics in Signal Processing}, vol.~12, no.~1, pp. 160--167, 2018.

\bibitem{riyaz2018deep}
S.~Riyaz, K.~Sankhe, S.~Ioannidis, and K.~Chowdhury, ``Deep learning convolutional neural networks for radio identification,'' \emph{IEEE Communications Magazine}, vol.~56, no.~9, pp. 146--152, 2018.

\bibitem{soltani2020more}
N.~Soltani, K.~Sankhe, J.~Dy, S.~Ioannidis, and K.~Chowdhury, ``More is better: Data augmentation for channel-resilient rf fingerprinting,'' \emph{IEEE Communications Magazine}, vol.~58, no.~10, pp. 66--72, 2020.

\bibitem{shen2021radio}
G.~Shen, J.~Zhang, A.~Marshall, L.~Peng, and X.~Wang, ``{Radio Frequency Fingerprint Identification for LoRa using Deep Learning},'' \emph{IEEE Journal on Selected Areas in Communications}, vol.~39, no.~8, pp. 2604--2616, 2021.

\bibitem{hanna2022wisig}
S.~Hanna, S.~Karunaratne, and D.~Cabric, ``Wisig: A large-scale wifi signal dataset for receiver and channel agnostic rf fingerprinting,'' \emph{IEEE Access}, vol.~10, pp. 22\,808--22\,818, 2022.

\bibitem{hamdaoui2022}
B.~Hamdaoui and A.~Elmaghbub, ``Deep-learning-based device fingerprinting for increased lora-iot security: Sensitivity to network deployment changes,'' \emph{IEEE Network}, vol.~36, no.~3, pp. 204--210, 2022.

\bibitem{elmaghbub2021lora}
A.~Elmaghbub and B.~Hamdaoui, ``Lora device fingerprinting in the wild: Disclosing rf data-driven fingerprint sensitivity to deployment variability,'' \emph{IEEE Access}, vol.~9, pp. 142\,893--142\,909, 2021.

\bibitem{al2021deeplora}
A.~Al-Shawabka, P.~Pietraski, S.~B. Pattar, F.~Restuccia, and T.~Melodia, ``Deeplora: Fingerprinting lora devices at scale through deep learning and data augmentation,'' in \emph{Proceedings of the Twenty-second International Symposium on Theory, Algorithmic Foundations, and Protocol Design for Mobile Networks and Mobile Computing}, 2021, pp. 251--260.

\bibitem{stallings2023}
W.~Stallings, \emph{{Cryptography and Network Security: Principles and Practice}}.\hskip 1em plus 0.5em minus 0.4em\relax Pearson Education, 2023.

\bibitem{ulversoy2010_comst}
T.~Ulversoy, ``{Software Defined Radio: Challenges and Opportunities},'' \emph{IEEE Communications Surveys \& Tutorials}, vol.~12, no.~4, pp. 531--550, 2010.

\bibitem{ettus}
{Ettus Research}, ``{USRP X310},'' \url{https://www.ettus.com/all-products/x310-kit/}, 2020, (Accessed: 2024-Mar-12).

\bibitem{uhd_man}
------, ``{USRP Hardware Driver and USRP Manual},'' \url{https://files.ettus.com/manual/page\_images.html}, 2020, (Accessed: 2024-Mar-12).

\bibitem{alhazbi2023ccnc}
S.~Alhazbi, S.~Sciancalepore, and G.~Oligeri, ``Bloodhound: Early detection and identification of jamming at the phy-layer,'' in \emph{IEEE Consumer Communications \& Networking Conference (CCNC)}, 2023, pp. 1033--1041.

\bibitem{sadighian2024ccnc}
A.~Sadighian, S.~Sciancalepore, and G.~Oligeri, ``{FadePrint}: Satellite spoofing detection via fading fingerprinting,'' in \emph{2024 IEEE 21th Consumer Communications \& Networking Conference (CCNC)}, 2024.

\bibitem{oligeri2023tifs}
G.~Oligeri, S.~Sciancalepore, S.~Raponi, and R.~D. Pietro, ``{PAST-AI}: Physical-layer authentication of satellite transmitters via deep learning,'' \emph{IEEE Transactions on Information Forensics and Security}, vol.~18, pp. 274--289, 2023.

\bibitem{irfan2023isncc}
M.~Irfan, A.~Omri, J.~H. Fernandez, S.~Sciancalepore, and G.~Oligeri, ``{Jamming Detection in Power Line Communications Leveraging Deep Learning Techniques},'' in \emph{Int. Symp. on Networks, Computers and Communications (ISNCC)}, 2023, pp. 1--6.

\bibitem{sciancalepore2023jamming}
S.~Sciancalepore, F.~Kusters, N.~K. Abdelhadi, and G.~Oligeri, ``{Jamming Detection in Low-BER Mobile Indoor Scenarios via Deep Learning},'' \emph{IEEE Internet of Things Journal}, vol.~11, no.~8, pp. 14\,682--14\,697, 2024.

\end{thebibliography}

\balance

\end{document}